\documentclass[aps, prx, reprint, superscriptaddress]{revtex4-2}

\usepackage{amsfonts}
\usepackage{amsmath}
\usepackage{amssymb, bm, mathrsfs}
\usepackage{graphicx, epstopdf, color}
\usepackage{soul}
\usepackage{braket}
\usepackage[version=4]{mhchem}
\usepackage{siunitx}
\usepackage{tabularx}
\usepackage[colorlinks=true,citecolor=blue]{hyperref}
\usepackage{array}
\usepackage{comment}

\newcolumntype{L}[1]{>{\raggedright\arraybackslash}p{#1}}

\newcommand{\fig}[1]{[Fig. \ref{#1}]}
\newcommand{\figsub}[2]{[Fig. \ref{#1}(#2)]}
\newcommand{\reftable}[1]{[Table. \ref{#1}]}

\begin{document}

\title{Chips in the \textit{Flatland}: 2D Semiconductors for Future Computing Electronics}

\author{Narin Trakarnvanich}
\affiliation{Science, Mathematics and Technology (SMT) Cluster, Singapore University of Technology and Design (SUTD), Singapore 487372.}

\author{Mitra Sanchali}
\affiliation{Science, Mathematics and Technology (SMT) Cluster, Singapore University of Technology and Design (SUTD), Singapore 487372.}

\author{Tong Su}
\affiliation{Science, Mathematics and Technology (SMT) Cluster, Singapore University of Technology and Design (SUTD), Singapore 487372.}

\author{Haiyu Meng}
\affiliation{School of Physics and Optoelectronics, Xiangtan University, Xiangtan 411105, China}

\author{\\Jing Lu}
\email{jinglu@pku.edu.cn}
\affiliation{State Key Laboratory for Mesoscopic Physics and School of Physics, Peking University, Beijing 100871, P. R. China}
\affiliation{Collaborative Innovation Center of Quantum Matter, Beijing 100871, P. R. China}
\affiliation{Beijing Key Laboratory for Magnetoelectric Materials and Devices (BKL-MEMD), Peking University, Beijing 100871, P. R. China}
\affiliation{Peking University Yangtze Delta Institute of Optoelectronics, Nantong 226010, P. R. China}
\affiliation{Key Laboratory for the Physics and Chemistry of Nanodevices, Peking University, Beijing 100871, P. R. China}
\affiliation{Beijing Key Laboratory of Quantum Devices, Peking University, Beijing 100871, P. R. China}

\author{Kah-Wee Ang}
\email{kahwee.ang@nus.edu.sg}
\affiliation{Department of Electrical \& Computer Engineering, National University of Singapore, Singapore 117575, Republic of Singapore}

\author{Lain-Jong Li}
\email{lance.li@nus.edu.sg}
\affiliation{Department of Materials Science \& Engineering, National University of Singapore, Singapore 117575, Republic of Singapore}

\author{Chit Siong Lau}
\email{aaron\_lau@a-star.edu.sg}
\affiliation{Quantum Innovation Centre (Q. InC), Agency for Science, Technology and Research (A*STAR), Singapore, 138635, Republic of Singapore}

\author{Yee Sin Ang}
\email{yeesin\_ang@sutd.edu.sg}
\affiliation{Science, Mathematics and Technology (SMT) Cluster, Singapore University of Technology and Design (SUTD), Singapore 487372, Republic of Singapore}

\begin{abstract}

As transistor scaling approaches its fundamental physical limits in the \emph{Angstrom era}, two-dimensional (2D) semiconductors have emerged as the promising channel material candidates for future computing. While the device physics of 2D semiconductors have been rigorously explored, translating these nanodevices into fully functional integrated circuits remains a largely uncharted frontier. This review bridges the gap between material- and device-centric breakthroughs and circuit-level chip design in 2D semiconductors -- a 'valley of death' that has so far prevented translation of high-performance individual transistors into functional chips. We track the evolution of 2D semiconductor field-effect transistors from basic Boolean logic families and standard cells to complex chip architectures, including recent milestones in RISC-V and monolithic CMOS microprocessors. Critically, we highlight the indispensable role of multiscale compact modeling, spanning semiclassical, quantum-hybrid and data-driven approaches, as the necessary link between device physics and the electronic design automation workflows for scalable chip development. By summarizing recent breakthroughs and identifying the bottlenecks in both "\emph{fab}" and "\emph{fabless}" trajectories of 2D semiconductors, this review shall provide insights that motivates the translation of proof-of-concept 2D transistors into fully functional computing chips, paving a way towards future Angstrom era computing technology empowered by 2D semiconductors.

\end{abstract}

\maketitle

\section{Introduction}

Silicon technology has dominated modern electronics for decades. Although the first field-effect transistor (FET) was implemented using germanium in 1947, the silicon metal-oxide-semiconductor field-effect transistor (MOSFET) gradually gained prominence due to its compatibility with large-scale circuit integration \cite{authEvolutionTransistorsHumble2023}. Silicon has now become the core material for modern computing chips. Over time, the silicon transistor has aggressively downscaled to nanometer dimensions, driven by the ever increasing demands on higher performance, lower energy consumption, and smaller device footprint. As a result, the transistor density and circuit complexity have scaled tremendously in the past decades. Microprocessors now are integrated with billions of transistors in highly complex architectures \cite{furberMicroprocessorsEnginesDigital2017, lundstromMooresLawJourney2022}. By 2025, a state-of-the-art central processing unit (CPU), such as Apple’s M3 Ultra, integrates more than 184 billion transistors. These chips are fabricated using the 3 nm (TSMC N3B) process technology \cite{Apple2025M3Ultra}, exemplifying how continuous transistor scaling remains the driving force behind next-generation computing and the rapid emergence of AI.

Since the early 2000s, silicon technology has shown signs of material limitations. According to Moore’s law, the number of transistors on a single chip was expected to double every two years \cite{serviceChipmakersLookMoores2018}. Despite this expectation the downscaling of transistors has slowed significantly due to fundamental challenges at the nanometer-scale where electron transport in ultra-thin silicon channels becomes highly sensitive to surface defects and charge traps that result in severe mobility degradation. Short-channel devices also suffer from increased tunneling effects, which contribute to higher static power consumption \cite{lundstromMooresLawJourney2022, shalfFutureComputingMoores2020, liHow2DSemiconductors2019, waltlPerspective2DIntegrated2022, wangTwodimensionalDevicesIntegration2022}. While FinFETs, gate-all-around FETs (GAAFETs), and complementary FETs (CFETs) have extended the silicon scaling roadmap, sustaining performance gains at sub-2-nm-equivalent nodes will likely require a channel material beyond silicon.

\begin{figure*}[t]
    \includegraphics[scale=0.34]{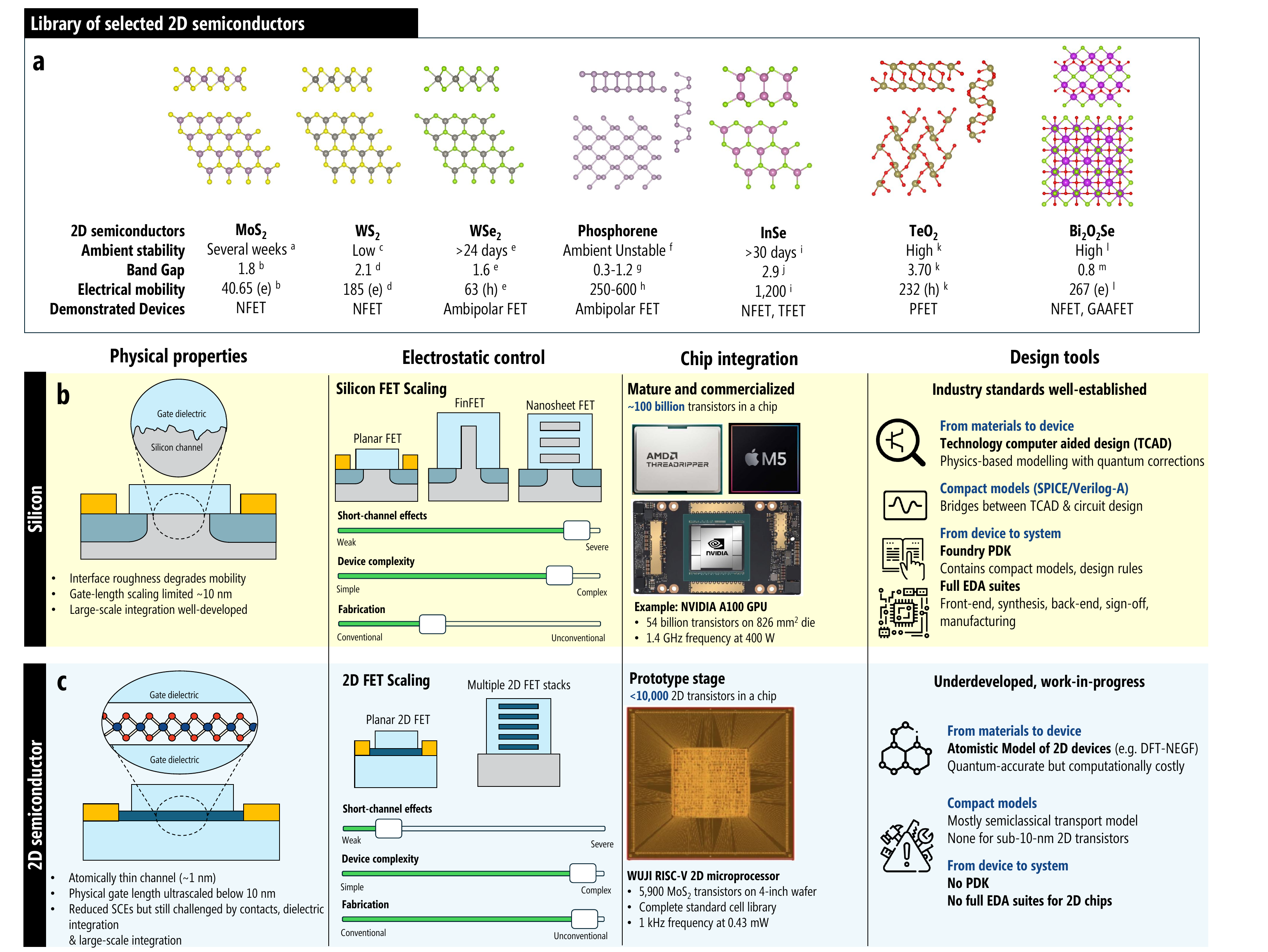}
    \caption{\textbf{Overview of 2D semiconductor technology and comparison with silicon technology} (a) A library of selected 2D semiconductors, including \ce{MoS2}, \ce{WS2}, \ce{WSe2}, phosphorene, \ce{InSe}, \ce{TeO2}, and \ce{Bi2O2Se}, is presented with information on their atomic structure, stability, band gap, electrical mobility [(e) for electron mobility and (h) for hole mobility], and implemented devices. (b) A qualitative comparison is provided between silicon and 2D semiconductor technologies in terms of physical properties, electrostatic control with available architectures, current chip integration, and design tools. While 2D semiconductors exhibit superior physical and electrical properties, further development is required in chip integration and design tool support to matched their device design and fabrication complexity. Reference: (Ref a.) \cite{chenDegradationBehaviorsMechanisms2018} (Ref b.) \cite{xia12inchGrowthUniform2023} (Ref c.) \cite{knoblochImprovingStabilityTwodimensional2022} (Ref d.) \cite{iqbalHighmobilityAirstableSinglelayer2015} (Ref e.) \cite{lanUncoveringDopingMechanism2025} (Ref f.) \cite{wangDegradationPhosphoreneAir2016} (Ref g.) \cite{heHighPerformanceBlackPhosphorus2019} (Ref h.) \cite{huangBlackPhosphorusElectronics2019} (Ref i.) \cite{jiangStableInSeTransistors2019} (Ref j.) \cite{hamerIndirectDirectGap2019} (Ref k.) \cite{zavabetiHighmobilityPtypeSemiconducting2021} (Ref l.) \cite{wangPolarityModulationCompositionally2025} (Ref m.) \cite{yuIntegrated2DMultifin2024}
    }
    \label{fig:vssilicon}
\end{figure*}

Two-dimensional (2D) semiconductors are compelling candidates for advancing transistors into the angstrom-scale technology nodes \cite{meenaSub5Nm2D2023}. Among the diverse family of 2D semiconductors, transition metal dichalcogenides (TMDs) monolayers of molybdenum disulfide (\ce{MoS2}) \cite{sebastianBenchmarkingMonolayerMoS22021, desaiMoS2Transistors1nanometer2016} (NMOS) and tungsten diselenide (\ce{WSe2}) \cite{chengWSe22DPtype2020} have been extensively studied for $n$-type (NMOS) and $p$-type MOSFETs (PMOS) applications, respectively \cite{wangThreedimensionalTransistorsIntegration2023}. With absence of surface dangling in their ultrathin morphology \cite{duttaElectronicProperties2D2024}, 2D semiconductors retain high electrical carrier mobility even at sub-1-nm channel thickness. Together with its high electrostatic gate control from ultrathin body, short-channel effects (SCE) can be strongly suppressed in 2D semiconductor as compared to bulk silicon which experienced severe mobility degradation and poor electrostatic gate control at nanometer thickness \cite{wangRoad2DSemiconductors2022, shenTrend2DTransistors2022}. 
Beyond \ce{MoS2} and \ce{WSe2}, a diverse family of 2D semiconductors such as tungsten disulfide (\ce{WS2}) \cite{realeHighMobilityHighOpticalQuality2017, sebastianBenchmarkingMonolayerMoS22021}, indium selenide (\ce{InSe}) \cite{daiPropertiesSynthesisDevice2022}, black phosphorus (\ce{BP}) \cite{liBlackPhosphorusFieldeffect2014} and MA$_2$Z$_4$ \cite{latychevskaiaNewFamilySeptuplelayer2024, thoCataloguingMoSi2N4WSi2N42023, zhouVanWaalsMoSi2N42025}, have also been demonstrated to exhibit good performance for FET applications.

The lab-based demonstration of high-performance 2D FETs has further motivated efforts toward large-scale, circuit-level implementation of 2D transistors into computing chips, as exemplified by recent proof-of-concept demonstrations of 1-bit microprocessors and RISC-V architectures \cite{aoRISCV32bitMicroprocessor2025}. However, integrating individual transistors into higher-level designs is a highly non-trivial task. Beyond achieving high-performance devices, successful chip integration requires overcoming material-and device-level bottlenecks such as the growth of high-quality and high-uniformity wafer-scale 2D semiconductors \cite{dongAirStableLargeArea2D2024, yinRecentProgressGrowth2025}, high-yield 2D circuit fabrication \cite{lanzaYieldVariabilityReliability2020}, reliable dielectric interfaces \cite{zhangDielectricIntegrationsAdvanced, linDielectricMaterialTechnologies2023, lauDielectricsTwoDimensionalTransitionMetal2023} and low-resistance electrical contacts \cite{choiRecentProgress1D2022, maVanWaalsContact2024}. 
Recent breakthrough in fabricating 6-inch MoS$_2$ wafer with electron mobility over 100 cm$^2$V$^{-1}$s$^{-1}$ \cite{liu2026kinetic} paves a solid ground towards the development of 2D computing chips. A deep understanding on the circuit behavior and signal interactions when many 2D devices are compactly interconnected on a single chip plays an equally important role. While 2D semiconductor research has achieved significant progress at the material and device levels, circuit-level design that includes compact modeling, standard cell development, and large-scale logic integration is still in its early stages. Advancing 2D semiconductors toward fully functional chips thus requires coordinated efforts in both device fabrication and circuit-level design methodologies.

Earlier reviews have comprehensively covered 2D semiconductors \cite{chhowallaTwodimensionalSemiconductorsTransistors2016} and individual device performance \cite{dasTransistorsBasedTwodimensional2021, liuTwodimensionalMaterialsNextgeneration2020}. Here, in contrast, we focus specifically on the under-reviewed transition from devices to chips, covering compact modeling, standard cell libraries, microprocessor architectures and EDA workflow gaps that determine whether device-level breakthroughs can translate into functional integrated circuits. We aim bridge the gap between the 2D semiconductor device and circuit design communities. Although both domains are essential in advancing 2D computing chip technology, they have traditionally progressed in different directions -- the device community focuses on materials, interfaces, and device physics, while the circuit community emphasizes logic design, standard-cell libraries, and system-level integration. Critically, the development of 2D chips is inherently multidisciplinary and requires close collaboration between these domains, where device-level physics informs circuit abstraction, and circuit requirements in turn guide device optimization and modeling. Here, a clear gap remains, most notably the absence of mature compact modeling frameworks for 2D semiconductors, thus creating a "\emph{valley of death}" between device-level research and circuit-level design. 

To this end, this Review shall provide a holistic overview of 2D semiconductor technology across multiple abstraction levels, from individual FETs to logic gates and microprocessors. We examine emerging design tools and compact modeling approaches that connect 2D device physics to circuit-level simulation and chip architecture. Finally, we discuss the key challenges and future prospects of 2D chip technologies. We hope that this article can serve as a timely bridge, introducing the state-of-the-art in 2D semiconductor devices to the circuit design community, while emphasizing the critical need for compact modeling and design frameworks to unlock the full potential of 2D semiconductors for future computing.

\begin{figure*}[t]
    \includegraphics[scale=0.505]{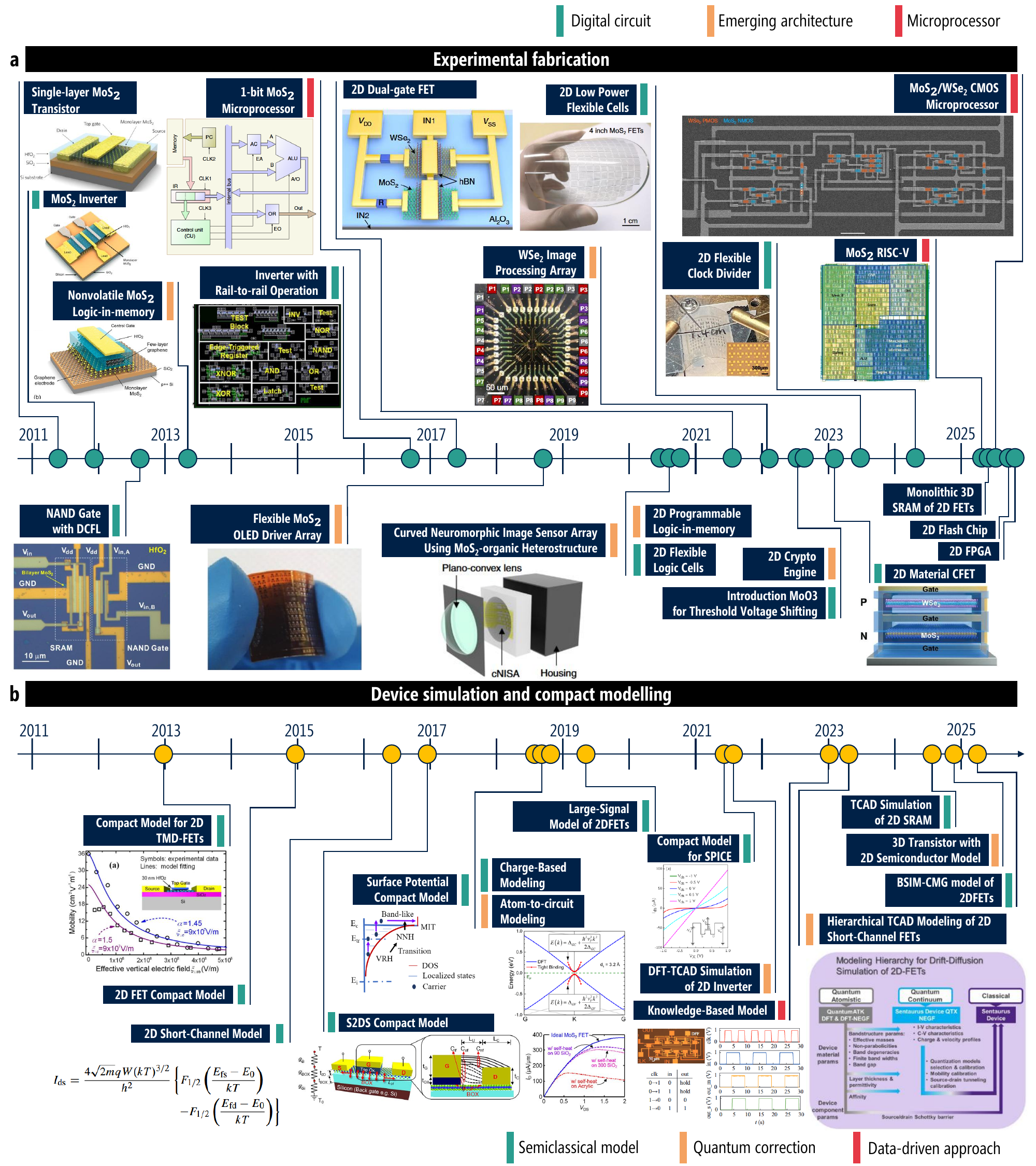}
    \caption{\textbf{Timeline of 2D semiconductors from materials to device and to integrated circuit design} (a) the timeline of experimental studies                  \cite{radisavljevicSinglelayerMoS2Transistors2011,
        radisavljevicIntegratedCircuitsLogic2011,
        wangIntegratedCircuitsBased2012,
        bertolazziNonvolatileMemoryCells2013,
        yuDesignModelingFabrication2016,
        wachterMicroprocessorBasedTwodimensional2017,
        wooLargeAreaCVDGrownMoS22018,
        choiCurvedNeuromorphicImage2020,
        liLargescaleFlexibleTransparent2020,
        migliatomaregaLogicinmemoryBasedAtomically2020,
        chenLogicGatesBased2021,
        zengApplicationspecificImageProcessing2022,
        doddaAllinoneBioinspiredLowpower2022,
        tianRailtoRailMoS2Inverters2022,
        liuLargeScaleUltrathinChannel2023,
        pengMediumscaleFlexibleIntegrated2024,
        aoRISCV32bitMicroprocessor2025,
        sadafEnablingStaticRandomaccess2025,
        ghoshComplementaryTwodimensionalMaterialbased2025,
        liuFullfeatured2DFlash2025}
      (b) In parallel to hardware development, compact modelling of 2D semiconductor transistors represent another emerging field that are of critical importance in driving forward the R\&D of 2D semiconductor chips. 
      \cite{jimenezDriftdiffusionModelSingle2012,caoCompactCurrentVoltage2014,taurShortChannelModel2D2016,suryavanshiS2DSPhysicsbasedCompact2016,wangSurfacePotentialBased2018,yadavChargeBasedModelingTransition2018,dasAtomtocircuitModelingApproach2018,pasadasLargesignalModel2DFETs2019,ahsanSPICECompactModel2021,rawatPerformanceProjection2D2021,qiKnowledgebasedNeuralNetwork2023,silvestriHierarchicalModelingTCAD2023,luProjectedPerformanceSi2024,palThreedimensionalTransistorsTwodimensional2024a,chenBSIMCompactModel2025}
    }
    \label{fig:timeline}
\end{figure*}

\section{A Hitchhiker's Guide on 2D semiconductors: From monolayers Towards chips}

\subsection{The amazing world of 2D semiconductors}
The discovery of graphene in 2004 \cite{novoselovElectricFieldEffect2004} ignited a revolution in solid-state electronics, giving rise to what is now often called the “flatland” of two-dimensional (2D) materials \cite{geimRiseGraphene2007, choiGrapheneSynthesisApplications2012, schwierzGrapheneTransistors2010, castronetoElectronicPropertiesGraphene2009, weissGrapheneEmergingElectronic2012}. With their atomically thin bodies, 2D materials offer extreme electrostatic control, mechanical flexibility, and seamless integration compatibility with diverse substrates, which are difficult to achieve in conventional bulk semiconductors. The isolation of graphene quickly inspired exploration of semiconducting and insulating analogues such as transition metal dichalcogenides (TMDCs) \cite{liuTwoDimensionalSemiconductorsMaterials2017, liuTwodimensionalTransistorsGraphene2018}, hexagonal boron nitride (hBN) \cite{mericGrapheneFieldEffectTransistors2013, zhangTwoDimensionalHexagonal2017, franklinNanomaterialsTransistorsHighperformance2015}, and beyond, expanding the 2D material family far beyond a single elemental crystal \cite{zhangHighthroughputComputationalScreening2019}.  

Following the discovery of graphene, transition metal dichalcogenides (TMDCs) have emerged as a leading class of next-generation 2D semiconductors. TMDCs adopt the chemical formula \(\ce{MX2}\), where M is a transition metal (such as Mo, W, Ti, or Nb) and X is a chalcogen (S, Se, or Te). Each monolayer consists of a transition-metal plane sandwiched between two chalcogen planes, forming a robust covalently bonded trilayer stabilized by van der Waals interactions between adjacent layers \cite{manzeli2DTransitionMetal2017, wangMoS2MaterialsPhysics2014, omarazzaroniConceptsDesignMaterials2022, wangThreedimensionalTransistorsIntegration2023}. Among them, molybdenum disulfide (\(\ce{MoS2}\)) has become the prototypical semiconducting TMDC and remains one of the most widely used 2D channel material for transistor and circuit demonstrations. Other noteworthy TMDC semiconductors include tungsten diselenide (\(\ce{WSe2}\)) \cite{allainElectronHoleMobilities2014, zengApplicationspecificImageProcessing2022}, tungsten disulfide (\(\ce{WS2}\)) \cite{realeHighMobilityHighOpticalQuality2017, sikhwangTransistorsChemicallySynthesized2012}, molybdenum ditelluride (\(\ce{MoTe2}\)) \cite{pezeshkiStaticDynamicPerformance2016}, and rhenium disulfide (\(\ce{ReS2}\)) \cite{kwonAll2DReS2Transistors2019, kimLowvoltageComplementaryInverters2017}. Collectively, these materials span a broad spectrum of bandgaps (1–2 eV) and carrier polarities, offering versatile platforms for complementary logic, photonics, and memory devices.

Beyond TMDC, a diverse range of other 2D semiconductors have been explored to address specific performance bottlenecks. For instance, InSe monolayer exhibits exceptionally large electron mobility and strong thickness-dependent optical response, making it promising for high-performance electronics and optoelectronics \cite{jiangBallisticTwodimensionalInSe2023, zhangApproachingIntrinsicThreshold2022, daiPropertiesSynthesisDevice2022}. BP has attracted much attention due to its anisotropic carrier transport and thickness-tunable direct bandgap (ranging from $\sim$0.3 eV in bulk to $\sim$2.0 eV in monolayer form) suitable for both high-performance transistors and broadband optoelectronic applications \cite{huangBlackPhosphorusElectronics2019}. The expanding 2D semiconductor library thus provides a rich palette for electronic device design.

Beyond their monolayer form, 2D materials can be vertically `Lego-stacked' to form van der Waals heterostructures (vdWHs), in which dissimilar layers are held together by weak interlayer forces without lattice matching constraints \cite{geimVanWaalsHeterostructures2013, castellanos-gomezVanWaalsHeterostructures2022}. vdWH engineering enables a large variety of hybrid materials with designer functionalities not found their standalone monolayers. The emergence of twistronics—where rotating adjacent 2D layers by a small \textit{magic angle}, inducing moire superlattices and correlated quantum phenomena. This structure further transform and expand the landscape of 2D physics and device engineering \cite{carrTwistronicsManipulatingElectronic2017, wangCorrelatedElectronicPhases2020, nimbalkarOpportunitiesChallengesTwisted2020, houStrainEngineeringTwisted2025, caoUnconventionalSuperconductivityMagicangle2018}. 
As famously remarked by Richard Feynman in his 1959 lecture “\textit{There’s Plenty of Room at the Bottom}”, the emergence of 2D materials proves that there is indeed plenty of room in the flatland: a landscape where electronic, optical, and quantum properties can be sculpted layer by layer, angle by angle and atom by atom for novel functional device applications.

\subsection{2D semiconductors for future CMOS: Beyond the silicon frontier}
\label{subsec:2D_supremacy}

2D semiconductors have emerged as promising candidates to extend the scaling trajectory of complementary metal–oxide–semiconductor (CMOS) technology \cite{waltlPerspective2DIntegrated2022}. As silicon transistors approach their physical and electrostatic limits at physical gate lengths near 10 nm \cite{IRDS2024IRDS}, continued miniaturization faces well-known obstacles including short-channel effects (SCE), increased leakage currents, and mobility degradation. In contrast, 2D semiconductors offer several intrinsic advantages that directly address these bottlenecks \cite{chhowallaTwodimensionalSemiconductorsTransistors2016}: (i) ultrathin body structure; (ii) atomically clean, dangling-bond–free surfaces; (iii) compatibility with vertical stacking for 3D integration; and (iv) suitability for back-end-of-line (BEOL) monolithic integration.

\noindent {\textit{(i) Ultrathin-body.}}
The atomically thin channels of 2D semiconductors enable near-ideal gate electrostatics. Unlike bulk, planar, or FinFET architectures where only part of the channel responds efficiently to gate control, the entire body of a 2D FET is within the electrostatic reach of the gate, enabling significantly reduced drain-induced barrier lowering (DIBL), minimal short-channel effects, and sharp subthreshold swing (SS). The channel lengths of 2D transistor can thus be \textit{ultrascaled} to just a few nanometers while maintaining low leakage and robust switching behavior \cite{radisavljevicSinglelayerMoS2Transistors2011, desaiMoS2Transistors1nanometer2016}.

\noindent {\textit{(ii) angling-bond-free surfaces.}}
 The chemically inert surfaces of 2D semiconductors suppress interface-trap densities and Coulomb scattering, thus leading to improved gate coupling, reduced variability, and enhanced mobility. With properly engineered contacts and high-quality dielectrics, 2D FETs can achieve steep subthreshold slopes and high carrier mobility, often surpassing ultra-thin-body silicon channels \cite{qinSteepSlopeField2024, yangSteepslopeVerticaltransportTransistors2024}. Achieving large-area uniformity further improve process reproducibility, reliability, and yield - key enablers for industrial-grade 2D chips manufacturing \cite{zavabetiTwoDimensionalMaterialsLargeAreas2020, lanzaYieldVariabilityReliability2020}.

\noindent {\textit{(iii) Vertical stacking for 3D integration.}}
The van der Waals nature of 2D materials naturally facilitates vertical device stacking, thus opening a pathway to monolithic three-dimensional (3D) integration. As van der Waals layers do not require lattice matching and exhibit minimal interdiffusion, they can be seamlessly combined into multilayer architectures \cite{kimVanWaalsLayer2023, jiangUltimateMonolithic3DIntegration2019, chenLargescaleGateallaroundMoS22026}. A compelling example is the complementary FET (CFET), where $n$-type and $p$-type 2D FETs, such as \ce{MoS2}/\ce{WSe2} or \ce{ReS2}/\ce{WSe2}, are vertically aligned \cite{liuLargeScaleUltrathinChannel2023, zhangSimulation2DReS22024}. Such stacking significantly shortens interconnect lengths, reduces parasitic capacitances and enhances area efficiency. These are key ingredients for continued system-level scaling as lateral scaling saturates \cite{zhangNewStructureTransistors2024}.

\noindent {\textit{(iv) Back-end-of-line integration.}}
Many 2D semiconductors can be synthesized or transferred at temperatures below 400 $^\circ$C, thus making them compatible with the BEOL processing requirements. This low thermal budget allows monolithic integration of 2D devices directly on top of completed Si CMOS wafers, enabling hybrid 2D/3D logic \cite{palThreedimensionalTransistorsTwodimensional2024a, wangThreedimensionalTransistorsIntegration2023, jayachandranThreedimensionalIntegrationTwodimensional2024, guoVanWaalsPolarityengineered2024, chowdhury3DIntegrationFunctionally2025, zhangNewStructureTransistors2024} and in-memory computing architectures \cite{yinEmerging2DMemory2021, migliatomaregaLogicinmemoryBasedAtomically2020}. Their mechanical flexibility and environmental stability also permit co-integration with diverse platforms such as silicon, III–V semiconductors, and ferroelectrics, thereby expanding the design space for heterogeneous systems \cite{quellmalzLargeareaIntegrationTwodimensional2021, huang2DSemiconductorsSpecific2022}.

Owing to these advantages, 2D semiconductors offer not only superb electrostatic control and ultrascaling potential at the device level, but also strong system-level benefits resulting from their atomic-scale morphology. These competitive strengths position 2D transistors as a promising enabler for future `angstrom-scale’ technology nodes, a difficult achievement for silicon alone \cite{yoonEnablingAngstromEra2025, liRecentExperimentalBreakthroughs2024}.

Finally, we note that for 2D semiconductors to become a viable channel material for future CMOS integration, what we term "2D supremacy" must be concretely demonstrated. We define 2D supremacy as the demonstration of 2D-semiconductor transistors meeting or exceeding the performance of silicon FinFET or GAAFET benchmarks on at least three of the following five metrics simultaneously, at a matched physical gate length below 10 nm: (i) on-current per unit width at matched off-current; (ii) subthreshold swing; (iii) drain-induced barrier lowering (DIBL); (iv) contact resistance; and (v) device-to-device variability over a statistically meaningful sample (e.g. $n \geq 100$). To date, no published demonstration has met this bar -- high-performance 2D devices in the sub-10-nm regime remain confined to single "hero" device demonstrations rather than statistically validated arrays -- highlighting that much of the device-level research landscape for 2D semiconductors remains to be charted.

\subsection{Fab and Fabless routes of 2D semiconductor: A journey from device to chip}

The development of 2D semiconductor chips has advanced along two complementary pathways: experimental fabrication (`fab') and computational, model-driven design (`fabless') \cite{liHow2DSemiconductors2019}. These parallel trajectories (summarized in \fig{fig:timeline}) jointly shape the current landscape of 2D device and circuit technologies. While fabrication efforts focus on improving material device performance, yield, and scalability, the fabless wing of 2D chip R\&D leverages modeling and simulation to predict device behavior, optimize performance, and guide circuit-level design. As experimental development is inherently time-and resource-intensive \cite{weiTwoDimensionalSemiconducting2022, liRecentExperimentalBreakthroughs2024}, predictive simulations and compact models have become indispensable for accelerating the device-to-circuit design loop in 2D electronics.

2D circuit demonstrations began shortly after the emergence of \ce{MoS2} transistors in the early 2010s (\figsub{fig:timeline}{a}). Foundational digital blocks, including inverters, NAND gates, and simple logic chains, were first implemented to assess device uniformity, switching behavior, and fabrication feasibility \cite{radisavljevicSinglelayerMoS2Transistors2011, radisavljevicIntegratedCircuitsLogic2011, wangIntegratedCircuitsBased2012, tianRailtoRailMoS2Inverters2022}. The introduction of field-gate FET (FGFET) architectures subsequently enabled early forms of 2D logic-in-memory \cite{bertolazziNonvolatileMemoryCells2013}. As device quality improved, circuit complexity increased rapidly: by 2017, proof-of-concept 1-bit microprocessors based entirely on 2D semiconductors were realized \cite{wachterMicroprocessorBasedTwodimensional2017}. More recently, complementary CMOS logic using mixed 2D channel materials has been demonstrated to achieve improved noise margins and power efficiency \cite{ghoshComplementaryTwodimensionalMaterialbased2025}. Concurrently, the application space has broadened, with demonstrations across neuromorphic architectures \cite{chenLogicGatesBased2021, migliatomaregaLowPowerArtificialNeural2022, doddaAllinoneBioinspiredLowpower2022}, edge-sensing and image-processing systems \cite{choiCurvedNeuromorphicImage2020}, and flexible or wearable electronics \cite{liLargescaleFlexibleTransparent2020, tangLowPowerFlexible2023, pengMediumscaleFlexibleIntegrated2024}.

In parallel with these experimental advances, the fabless research pathway has focused on developing compact models and simulation frameworks capable of predicting 2D device behavior with high accuracy (\figsub{fig:timeline}{b}). Compact models enable fast circuit-level simulation, which is essential for scaling up 2D chips from individual transistors to complex architectures. Most reported models to date are rooted in semiclassical drift–diffusion transport, which provides computationally efficient descriptions but can miss key quantum-mechanical phenomena relevant at sub-10-nm dimensions \cite{knoblochModeling2DMaterialBased2023}. The difficulty of extending computational method from 3D to 2D lies on atomically thin and delicate defect control challenge from 2D nature. To address this limitation, hierarchical quantum-to-circuit modeling frameworks have been developed, combining density functional theory (DFT) at the material level with non-equilibrium Green’s function (NEGF) calculations at the device level to generate compact representations for circuit simulation \cite{dasAtomtocircuitModelingApproach2018, silvestriHierarchicalModelingTCAD2023, palThreedimensionalTransistorsTwodimensional2024a}. Beyond physics-based models, emerging data-driven approaches leverage machine learning to learn current-voltage relationships or extract compact-model parameters directly from experimental or simulated datasets, providing an alternative route toward rapid and scalable model generation \cite{qiKnowledgebasedNeuralNetwork2023}.

Complementary to each other, the `fab' and fabless’ timelines together illustrates a rapidly evolving ecosystem, where fabrication efforts establish the experimental feasibility of 2D transistors and integrated circuits, while simulations and compact modeling accelerate device optimization, design exploration, and system-level integration. Critically, the convergence of `fab' and fabless’ tools is essential for advancing from isolated device demonstrations to fully functional 2D chips. We thus anticipate increasingly close gap between fab and fabless pathways to define the next key frontier in propelling 2D semiconductors from materials to chips.
 
\section{2D transistor circuit design}

In this section, we review experimental studies on 2D semiconductor circuits, beginning with the fundamental logic building blocks and progressing toward complex microprocessors. The foundational concept of digital circuit is introduced through a discussion of simple inverters. Next, We review major logic families employed in 2D digital circuits, their advantages and limitations, and how they shape circuit performance. The development of logic cells and their integration into larger computational modules are examined, followed by a survey of 2D memory circuits and emerging computing paradigms that extend beyond conventional digital logic.

\begin{figure*}[t]
    \includegraphics[scale=0.425]{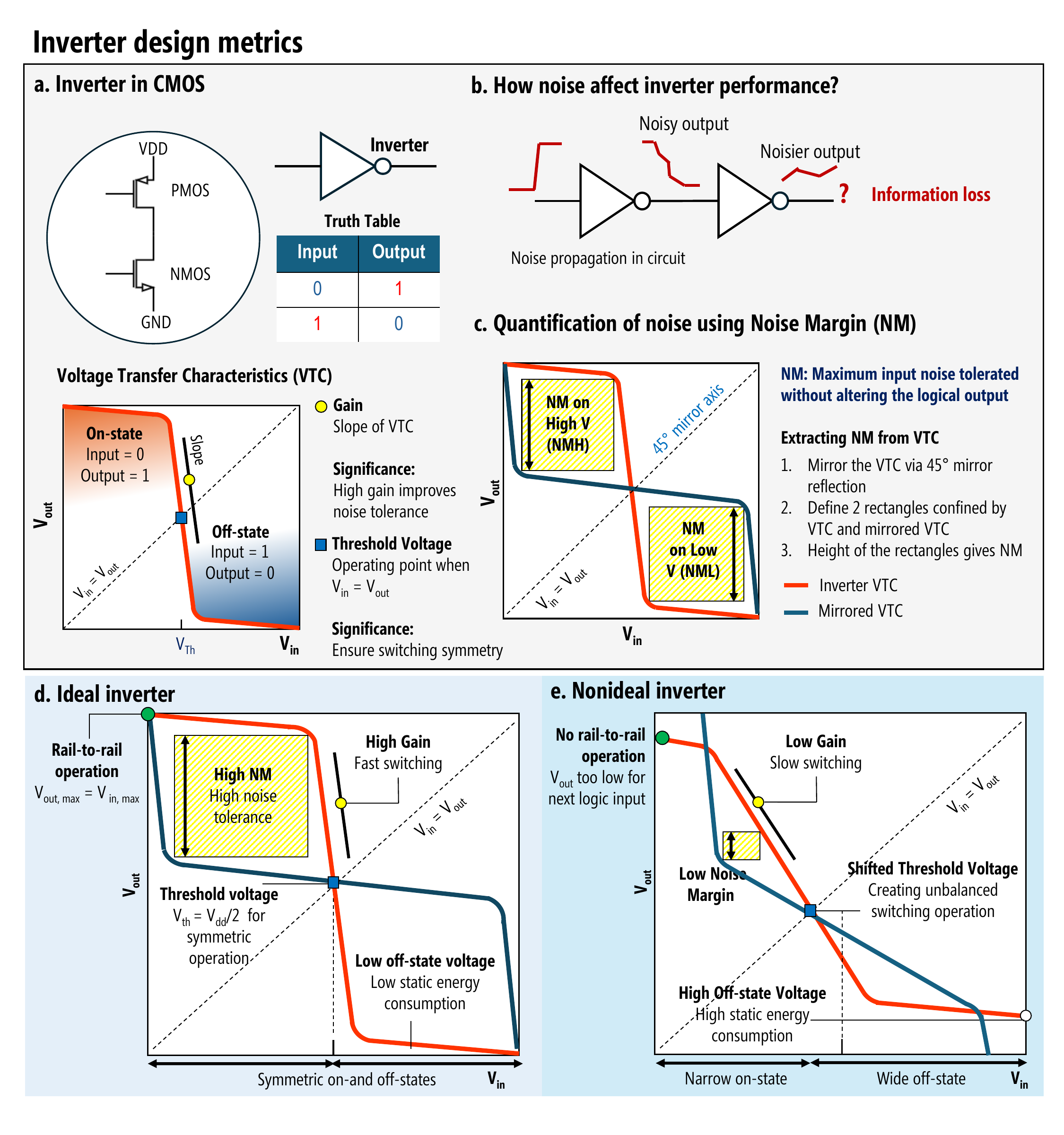}
    \caption{\textbf{Performance characteristics of 2D inverters.} (a) CMOS inverter consist of a pull-up PMOS and a pull-down NMOS transistor. When the input voltage is low, the PMOS drives the output to $V_{DD}$; when the input is high, the NMOS pulls the output to ground. (b) Noise propagation in cascaded logic stages. (c) Extraction of noise margin from the voltage transfer characteristic (VTC), showing the noise margin low (NML) and noise margin high (NMH). Comparison between an (d) ideal and (e) nonideal inverter. A high gain (steep slope) in the transition region indicates sharp switching and strong signal regeneration. A mid-range threshold voltage supports symmetric switching. full rail-to-rail output ensures compatibility across logic stages and reduces static power consumption.}
    \label{fig:inverter}
\end{figure*}

\subsection{Inverter: building block of digital circuits}
\label{sec:inverter_primer}

A inverter is the fundamental building block of digital logic and the simplest circuit used to characterize the performance overall digital circuit. Inverter outputs the logical complement of its input. In CMOS implementations, an NMOS transistor serves as the pull-down transistor (drive device), driving the output toward ground when the input is high, while a PMOS pull-up transistor (load device) restores the output to $V_{DD}$ when the input is low \cite{uyemuraCMOSInverterAnalysis2001}. This complementary action produces the characteristic voltage transfer curve (VTC), in which the output remains high until the input approaches the switching threshold, after which the output transitions sharply toward zero \figsub{fig:inverter}{a}. 

A key metric for evaluating inverter quality is the noise margin, which quantifies the tolerance of the inverter to disturbances when its output drives subsequent inverter stages \figsub{fig:inverter}{b}. High noise margins are essential as digital circuits consist of cascaded logic cells whose outputs accumulate noise through successive operations. Two values are typically extracted, the noise margin high (NMH) and noise margin low (NML), which are defined by the horizontal separation between the VTC and its 45-degree mirror \figsub{fig:inverter}{c}. Here, NML and NMH quantify the maximum noise voltage that can be tolerated at the low and high logic levels, respectively, without causing a logic error. Larger noise margins indicate more reliable signal restoration, and thus enabling deeper logic pipelines with more complex functionality.

Beyond noise margin, inverters are further assessed for their gain, threshold voltage, output voltage magnitude and compatibility with rail-to-rail operation \figsub{fig:inverter}{d, e}. These parameters determine how closely an inverter approaches ideal behavior and anticipate its performance. Gain is the steepness of the voltage transfer curve in the transition region and indicates how effectively an inverter amplifies small input variations into full logic-level swings. High gain enables sharp and reliable switching. The threshold voltage ($V_\text{th}$) marks the input voltage at which the inverter switches its output state. A mid-range $V_\text{th}$ supports symmetric rising and falling transitions. Low off-state output voltage further minimizes static power consumption. Finally, the rail-to-rail operation, in which the inverter delivers output voltages spanning the full input range, ensures compatibility across logic stages and is crucial for reliable signal restoration in 2D digital circuits \cite{tianRailtoRailMoS2Inverters2022}.

\begin{figure*}[t]
    \includegraphics[scale=0.47]{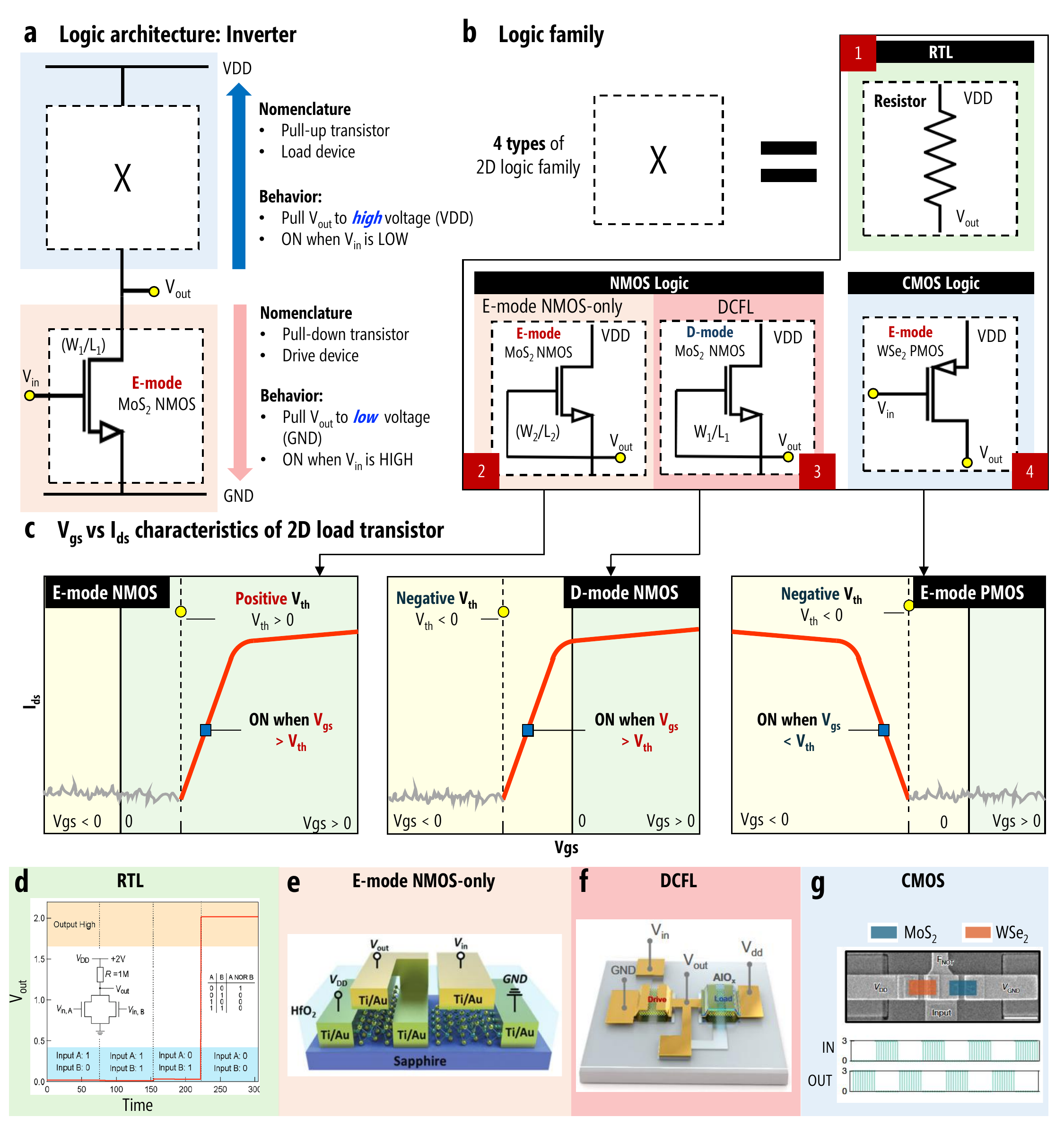}
    \caption{\textbf{2D Logic architectures, families, and load transistor choices} (a) Logic architecture for 2D inverter (b) Logic families with corresponding load devices (c) $V_\text{gs}$-$I_\text{ds}$ characteristic plots for 2D load transistors: E-mode NMOS, D-mode NMOS, and E-mode PMOS (d, e, f, g) The circuit implementation example of logic families: (d) RTL \cite{radisavljevicIntegratedCircuitsLogic2011}, (e) E-mode NMOS-only \cite{wangElectronicDevicesCircuits2019}, (f) DCFL \cite{pengMediumscaleFlexibleIntegrated2024}, (g) CMOS logic \cite{ghoshComplementaryTwodimensionalMaterialbased2025} 
    }
    \label{fig:transistor_mode}
\end{figure*}

\subsection{Transistor mode of operations on 2D inverter implementation}

Besides CMOS logic, digital circuit can be implemented with many logic families based on their device choice \figsub{fig:transistor_mode}{a, b}. 2D digital circuit design begins with the design of an inverter because it is the most basic building block. The inverter can be implemented with a combination of pull-up transistor (load device) and pull-down (drive device) transistor that connect output logic to the $V_\text{dd}$ and GND, respectively \cite{uyemuraCMOSInverterAnalysis2001, rabaeyDigitalIntegratedCircuits2003} \figsub{fig:transistor_mode}{a}. The inverter prefers a load device that turns ON only when $V_\text{in}$ is low, producing a high $V_\text{out}$. Conversely, the drive device turn ON when $V_\text{in}$ is high to pull $V_\text{out}$ down to a low voltage. Because both transistor has $V_\text{out}$ on drain terminal and pulling voltage on source terminal, they achieve preferred behavior for the inverter.

In inverter design, it is of key importance to consider two types of transistor operation modes: enhancement-mode (E-mode) and depletion-mode (D-mode). An E-mode transistor does not form a conducting channel when $V_\text{gs} = 0$~V; thus, it is a normally-OFF transistor. In contrast, a D-mode transistor is a normally-ON device that has a formed channel even at $V_\text{gs} = 0$~V. The ON-state region of a D-mode transistor overlaps with $V_\text{gs} = 0$~V~\figsub{fig:transistor_mode}{c}. This variety of operation modes, along with the two types of transistors (N-type and P-type), creates four distinct classes of transistors with designated applications. Three out of four, E-mode NMOS, D-mode NMOS, and E-mode PMOS, have been implemented in various 2D circuit experimental studies as load devices in 2D digital circuits. These configuration choices determine the logic families of the 2D circuit.

Based on these behaviors, an inverter prefer E-mode NMOS as the drive device., with the source connected to GND, the drain to $V_\text{out}$, and the gate to $V_\text{in}$. The E-mode NMOS turns ON only when $V_\text{gs} > V_\text{th}$, where $V_\text{th}$ is a positive threshold voltage~\figsub{fig:transistor_mode}{c}. This structure ensures that the output voltage $V_\text{out} = 0$ V when $V_\text{in}$ is HIGH. In the case of 2D inverter, the preferred devices are E-mode \ce{MoS2} NMOS transistors. 

Unlike the drive device, four possible load device configurations have been tested in 2D digital circuits, each corresponding to a different logic family \figsub{fig:transistor_mode}{b}. The simplest load device is a resistor, which forms a resistor-transistor logic (RTL) family. Although RTL has a simple architecture, the load resistor does not completely stop pulling up the voltage at HIGH $V_\text{in}$. Therefore, transistor-based load devices are preferred for full voltage control.

Early efforts in 2D semiconductor circuits typically employed NMOS-only (NMOS logic) architectures due to the lack of high-performance 2D PMOS transistors. These designs used \ce{MoS2} NMOS devices as load device, with both the drain and gate connected to $V_\text{out}$ to create a rectifying behavior similar to a diode. This architecture was used in early digital circuit designs before CMOS integration and is still used in GaN-based devices \cite{jiangThresholdVoltageModulation2024, panMonolithicLogicUnits2023} \figsub{fig:transistor_mode}{b}. Although NMOS logic can implement digital circuits similar to CMOS, it struggles to achieve a mid-range threshold inverter ($V_\text{th} = V_\text{dd}/2$) for large-scale integration when both load and drive devices are identical \cite{wachterMicroprocessorBasedTwodimensional2017}. Shifting the $V_\text{th}$ of the load device is necessary to adjust the inverter's threshold. Techniques such as modifying the channel width and length of E-mode NMOS devices have been studied to fabricate \ce{MoS2} NMOS transistors with varied threshold voltages \cite{wachterMicroprocessorBasedTwodimensional2017, wangElectronicDevicesCircuits2019}. These studies prove that digital circuits with only E-mode NMOS can achieve mid-range $V_\text{th}$ inverters without introducing new fabrication processes or materials. Another approach is using the negative $V_\text{th}$ of D-mode \ce{MoS2} NMOS to implement direct-coupled FET logic (DCFL). The DCFL structure is similar to the E-mode NMOS-only circuit family but shifts the $V_\text{th}$ through transistor material design rather than device dimensions, enabling digital circuits with identical transistor sizes \cite{aoRISCV32bitMicroprocessor2025, tianRailtoRailMoS2Inverters2022, pengMediumscaleFlexibleIntegrated2024, wangIntegratedCircuitsBased2012}.

The development of CMOS integration has been driven by advances in E-mode p-type 2D transistors, primarily \ce{WSe2}, paired with E-mode \ce{MoS2} NMOS devices to form CMOS logic \cite{ghoshComplementaryTwodimensionalMaterialbased2025}. CMOS uses the switching characteristics of E-mode PMOS, which turns ON when $V_\text{gs}$ is highly negative \figsub{fig:transistor_mode}{c}. At low $V_\text{in}$, the gate voltage is much lower than the source voltage (connected to $V_\text{dd}$) \figsub{fig:transistor_mode}{b}, resulting in a pull-up behavior. The CMOS architecture provides several advantages for 2D digital circuits, which are discussed in the following section.

\subsection{Logic families}

Each logic family offers distinct trade-offs in power consumption, switching speed, noise margin, design complexity, and scalability. These factors collectively shape the performance of 2D digital circuits. A common way to compare these logic families is through the performance of their fundamental building block, i.e. the \textit{inverter}. High-quality inverters exhibiting high noise margins, rail-to-rail swings, and low leakage current are essential for building robust, large-scale digital systems. To pursue these characteristics, the transistor with high ON/OFF ratio and sharp subthreshold slope is necessary \cite{choiSteepSwitchingWSe22022, duLowpowerconsumptionCMOSInverter2021}. Beyond device-level requirements, this effort also has motivated the transition from simpler RTL and NMOS logic toward CMOS logic, which offers superior noise immunity and energy efficiency.

In the following subsections, we examine each logic family in turn, beginning with 2D RTL circuits as the earliest and simplest demonstration of 2D digital logic.

\subsubsection{2D RTL circuit}

RTL is the simplest logic family, using a resistor as the load device. Its straightforward structure enabled early demonstrations of 2D digital circuits, but its performance is fundamentally limited by poor noise margins and substantial static power consumption. An RTL NOR gate with a 1~\si{\mega\ohm} load resistor can generate correct logic outputs \cite{radisavljevicIntegratedCircuitsLogic2011} \figsub{fig:transistor_mode}{d}, yet the resistive load produces significant leakage whenever at least one input is high. Due to these drawbacks, RTL remains confined to proof-of-concept demonstrations, and more advanced logic families are needed to reduce leakage and support scalable, energy-efficient 2D digital systems.

\subsubsection{2D NMOS logic families: DCFL and E-mode NMOS-only logic}

While CMOS logic is the dominant choice in silicon technology, its requirement for complementary 2D semiconductor types poses a major challenge for early 2D transistor design , which were largely limited to $n$-type behavior, most notably \ce{MoS2}. As a result, NMOS logic families such as DCFL and E-mode NMOS-only logic emerged as the most practical design options for 2D circuits in the 2010s \cite{kongRecentProgressesNMOS2021}. Both families can achieve rail-to-rail inverters with high noise margins and fast switching without relying on $p$-type transistors \cite{wachterMicroprocessorBasedTwodimensional2017, wangElectronicDevicesCircuits2019, pengMediumscaleFlexibleIntegrated2024}.

DCFL and E-mode NMOS-only logic differ in their load structures and resulting trade-offs. DCFL employs a D-mode transistor as an active pull-up, enabling full output voltage swing. In contrast, E-mode NMOS-only logic uses an E-mode transistor as the load, which restricts output swing; to compensate, transistor sizing is adjusted so that the load device has a larger width-to-length ratio than the drive transistor \cite{wangElectronicDevicesCircuits2019} \figsub{fig:transistor_mode}{e}. For multi-input gates, the drive transistors must also be widened to maintain sufficient pull-down strength \cite{wachterMicroprocessorBasedTwodimensional2017}. Consequently, DCFL tends to offer better area efficiency but requires additional steps or material engineering to realize D-mode transistors.

Multiple fabrication strategies have been developed for D-mode NMOS transistors. Gate-work-function engineering using different gate metals, such as Al/Pd \cite{wangIntegratedCircuitsBased2012} and Al/Au \cite{aoRISCV32bitMicroprocessor2025}, for load and drive devices can tune the threshold voltage. Threshold shifts can also be introduced via oxide capping layers such as \ce{AlO_x} \cite{pengMediumscaleFlexibleIntegrated2024} \figsub{fig:transistor_mode}{f}. Alternatively, ozone treatment of bilayer \ce{MoS2} forms a \ce{MoS2}/\ce{MoO3} heterostructure that yields a positive threshold shift \cite{tianRailtoRailMoS2Inverters2022}. These approaches enable rail-to-rail inverter performance without introducing new 2D semiconductors into the fabrication flow.

Despite the success of NMOS logic families (DCFL and E-mode NMOS-only logic), they encounter intrinsic limitations when scaling to multiple-input and more complex logic cells. As the number of inputs increases, the corresponding number of drive transistors also increases, which leads to signal level degradation. In these logic families, the pull-up device operates as a weak load, making it insufficient to support high fan-out circuits. Complex logic cells such as OR-AND-INVERT (OAI) or AND-OR-INVERT (AOI) further compound this issue, as multiple series and parallel paths increase network resistance and reduce noise margins. Indeed, DCFL implementations of higher-fan-in gates including NOR4, NAND4, AOI33, AOI22, OAI33, OAI221, and OAI222 show noise margins too low for reliable microprocessor operation \cite{aoRISCV32bitMicroprocessor2025}. These limitations motivate the transition toward CMOS logic family, which inherently provides stronger drive capability, improved noise margins, and lower static power consumption.

\subsubsection{2D CMOS circuit}

The emergence of two-dimensional (2D) $p$-type semiconductors, such as \ce{WSe2} \cite{chengWSe22DPtype2020, allainElectronHoleMobilities2014}, has enabled the development of CMOS logic family using 2D semiconductors. CMOS logic employs NMOS devices for pull-down operation and PMOS devices for pull-up operation, offering inherently low static power consumption, strong noise margins, and stable rail-to-rail output. With the availability of both carrier polarities in 2D materials, CMOS represents the most energy-efficient and scalable logic family for large-scale 2D digital systems.

Heterogeneous CMOS integration using \ce{MoS2} NMOS and \ce{WSe2} PMOS devices has achieved ultra-low static power consumption, down to 16 pW at $V_{DD}=3$ V \cite{ghoshComplementaryTwodimensionalMaterialbased2025} \figsub{fig:transistor_mode}{g}, with switching energies on the order of 100pJ per transition. By contrast, representative NMOS logic circuits report static power levels around \SI{1.4}{\micro\watt} per gate \cite{wachterMicroprocessorBasedTwodimensional2017}. These comparisons highlight the significant energy-efficiency gains offered by complementary architectures. Beyond planar designs, 2D CMOS is also being explored within advanced three-dimensional configurations such as the complementary FET (CFET), where vertically stacked $n$-and $p$-type channels enable further improvements in area efficiency and interconnect scaling \cite{liuLargeScaleUltrathinChannel2023}. Recently, a flexible monolithic 3D structure based on 2D CMOS integration with  \ce{MoS2} and \ce{WSe2} inks has been demonstrated as inverter, NAND, and NOR gates \cite{zouFlexibleMonolithic3D2025}. As fabrication techniques for 2D $p$-type semiconductors continue to mature  \cite{chengWSe22DPtype2020, eftekhariTungstenDichalcogenidesWS22017}, CMOS logic is poised to become the dominant architecture for high-performance and energy-efficient 2D integrated circuits.

\begin{figure*}[t]
    \includegraphics[scale=0.4658]{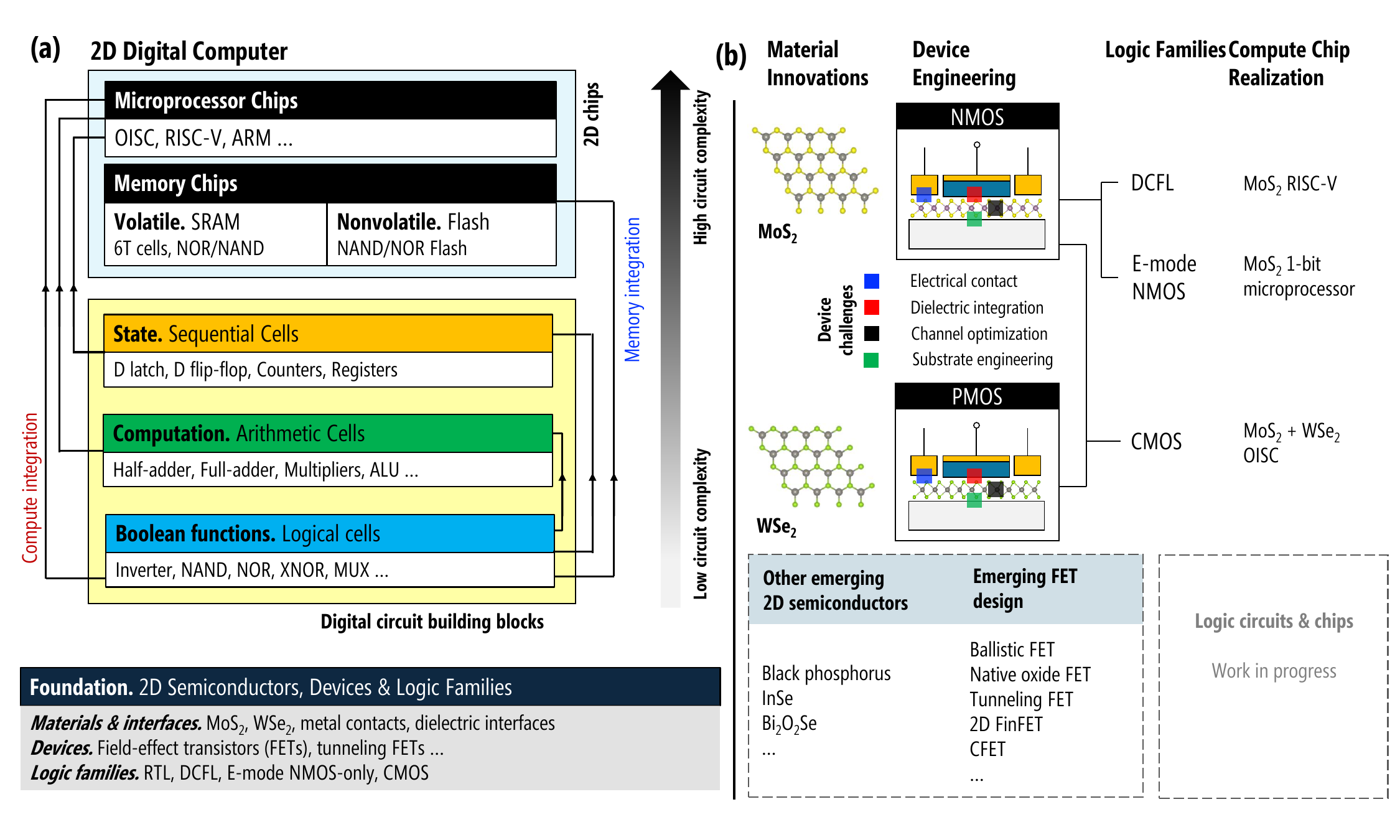}
    \caption{\textbf{Hierarchical landscape of 2D digital circuit and compute-chip development.} (a) The foundational layer consists of 2D semiconductors, device structures and logic families (RTL, DCFL, E-mode NMOS-only and CMOS). These elements support increasingly complex digital building blocks: Boolean logic cells, arithmetic cells and sequential/state elements. Integration of these blocks enables memory circuits and ultimately microprocessor-class architectures. (b) Pathways from materials and devices to logic and compute architectures. 2D CMOS technology is primarily driven by material innovations in\ce{MoS2} and \ce{WSe2}. At the device level, contacts, dielectric integration, channel optimization and substrate compatibility form the key enablers towards translation into different logic families—DCFL, E-mode NMOS, and CMOS. \ce{MoS2} RISC-V processors and \ce{MoS2}/\ce{WSe2} CMOS-based OISC architectures have been recently demonstrated. Beyond \ce{MoS2} and \ce{WSe2}, other 2D semiconductors and field-effect transistor (FET) design are also emerging, thus offering alternative pathways to achieve 2D chips. 
    }
    \label{fig:hierachy}
\end{figure*}

\subsection{Hierarchy of 2D Digital Circuit Blocks Toward Full-Function Chips}

The successful realization of all four layers using 2D semiconductors (see Fig.~\ref{fig:hierachy} for a schematics showing the hierarchy of digital design) demonstrates a rapidly advancing ecosystem in which device-level innovations increasingly translate into system-level functionality. Ongoing intensively over the past decades, 2D semiconductors have undergone a maturing design stack: (i) logic gates establish the basis for Boolean computation; (ii) arithmetic cells build upon them to enable numerical processing; (iii) sequential cells introduce timing and state retention for control logic and datapath coordination; (iv) memory circuits provide the storage backbone necessary for both fast on-chip access and long-term data retention and (v) full-functional 2D-semiconductor-based microprocessors and heterogeneous computing platforms. Below, we review these four layers of 2D digital circuit hierarchy, which shall form the backbone of a full-function micropocessor and computer enabled by 2D semiconductors.

\begin{figure*}[t]
    \includegraphics[scale=0.225]{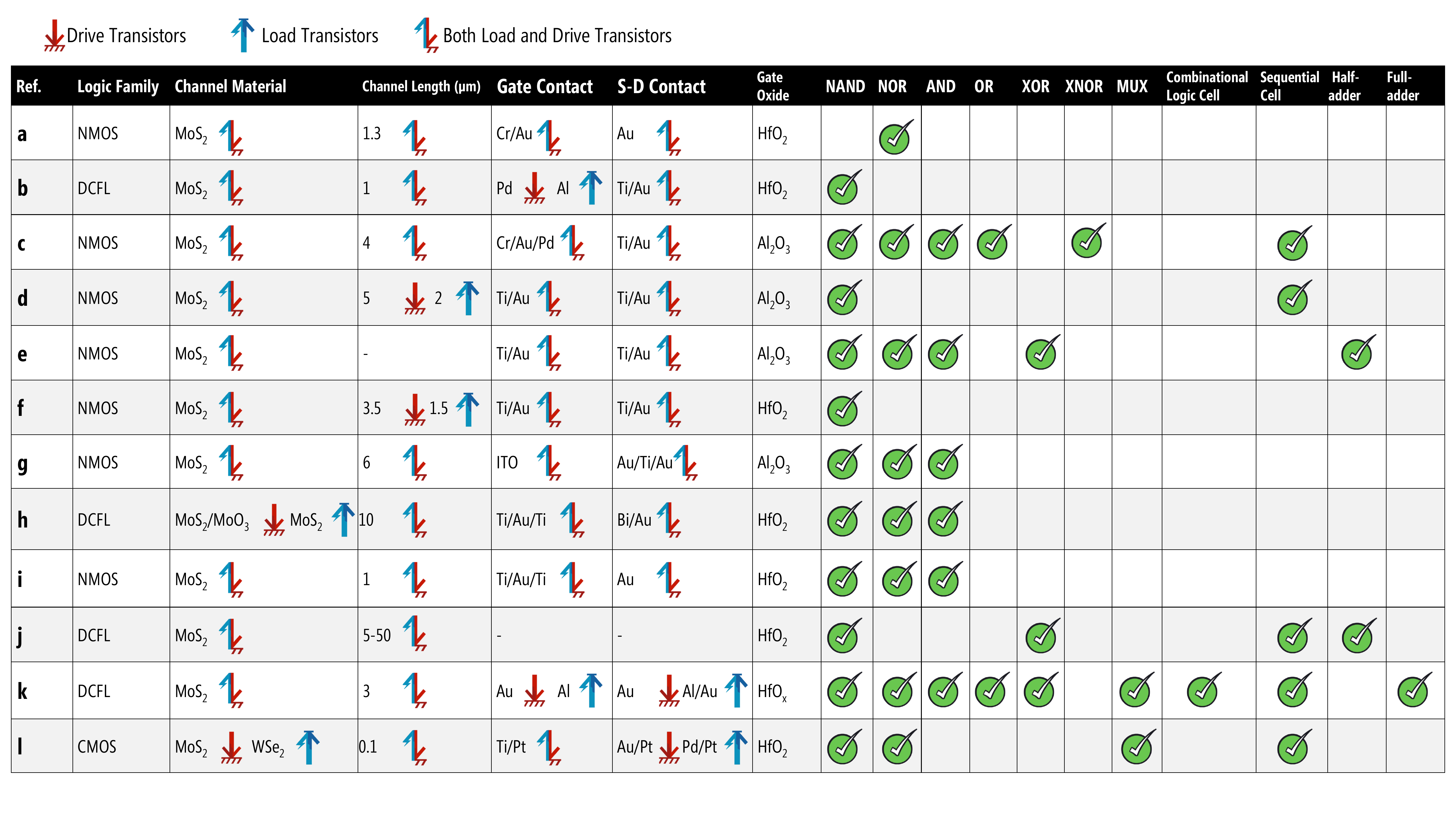}
    \caption{\textbf{2D transistors as logic cells.} Transistor specifications and their implementation as digital circuit logic cells. Drive and load transistors are transistor that have responsibility to pull-down and pull-up output voltage, respectively. The drive transistors connect between output and ground, while the load transistor connect between supply voltage and output. The circuit implementation of these transistors depends on logic family, intended logic operation, and circuit design. Combinational logic cells are a class of logic cells that include AOI and OAI gates. References:
    (Ref. a) \cite{radisavljevicIntegratedCircuitsLogic2011} 
    (Ref. b) \cite{wangIntegratedCircuitsBased2012} 
    (Ref. c) \cite{yuDesignModelingFabrication2016} 
    (Ref. d) \cite{wachterMicroprocessorBasedTwodimensional2017} 
    (Ref. e) \cite{linSolutionprocessable2DSemiconductors2018} 
    (Ref. f) \cite{wangElectronicDevicesCircuits2019} 
    (Ref. g) \cite{liLargescaleFlexibleTransparent2020}
    (Ref. h) \cite{tianRailtoRailMoS2Inverters2022}
    (Ref. i) \cite{tangLowPowerFlexible2023}
    (Ref. j) \cite{pengMediumscaleFlexibleIntegrated2024}
    (Ref. k) \cite{aoRISCV32bitMicroprocessor2025}
    (Ref. l) \cite{ghoshComplementaryTwodimensionalMaterialbased2025}
    }
    \label{fig:celltable}
\end{figure*}

\subsection{2D logic gates}

The establishment of 2D logic families enables the development of standard cells. 
Standard cells are the reusable building blocks of modern digital circuit design which allows large-scale integration by providing well-characterized logic elements with predictable electrical behavior \cite{zhouSurveyStandardCell2025}. The use of standard cell libraries reduces design complexity and improves the scalability as well as systematic verification and optimization within the electronic design automation (EDA) flows. A reliable standard-cell library is therefore essential for consistent circuit performance across different designs and fabrication processes, thus ultimately shortening design cycles and minimizing functional or timing errors.

2D semiconductor research has increasingly shifted toward the experimental realization of logic cells suitable for integration into microprocessors and other complex systems \cite{wachterMicroprocessorBasedTwodimensional2017, aoRISCV32bitMicroprocessor2025, ghoshComplementaryTwodimensionalMaterialbased2025}. Experimental 2D logic cells span a broad range, from basic inverters to multi-input logic gates and sequential elements such as flip-flops \fig{fig:celltable} \cite{pengMediumscaleFlexibleIntegrated2024, aoRISCV32bitMicroprocessor2025, ghoshComplementaryTwodimensionalMaterialbased2025}. These cells can broadly be classified into three categories: logic gate cells, arithmetic cells, and sequential cells. Each category is indispensable for enabling full microprocessor functionality. Recent developments in 2D logic gates will be reviewed below, beginning with a discussion on inverters and ring oscillators.

\begin{figure*}[t]
    \includegraphics[width=\textwidth]{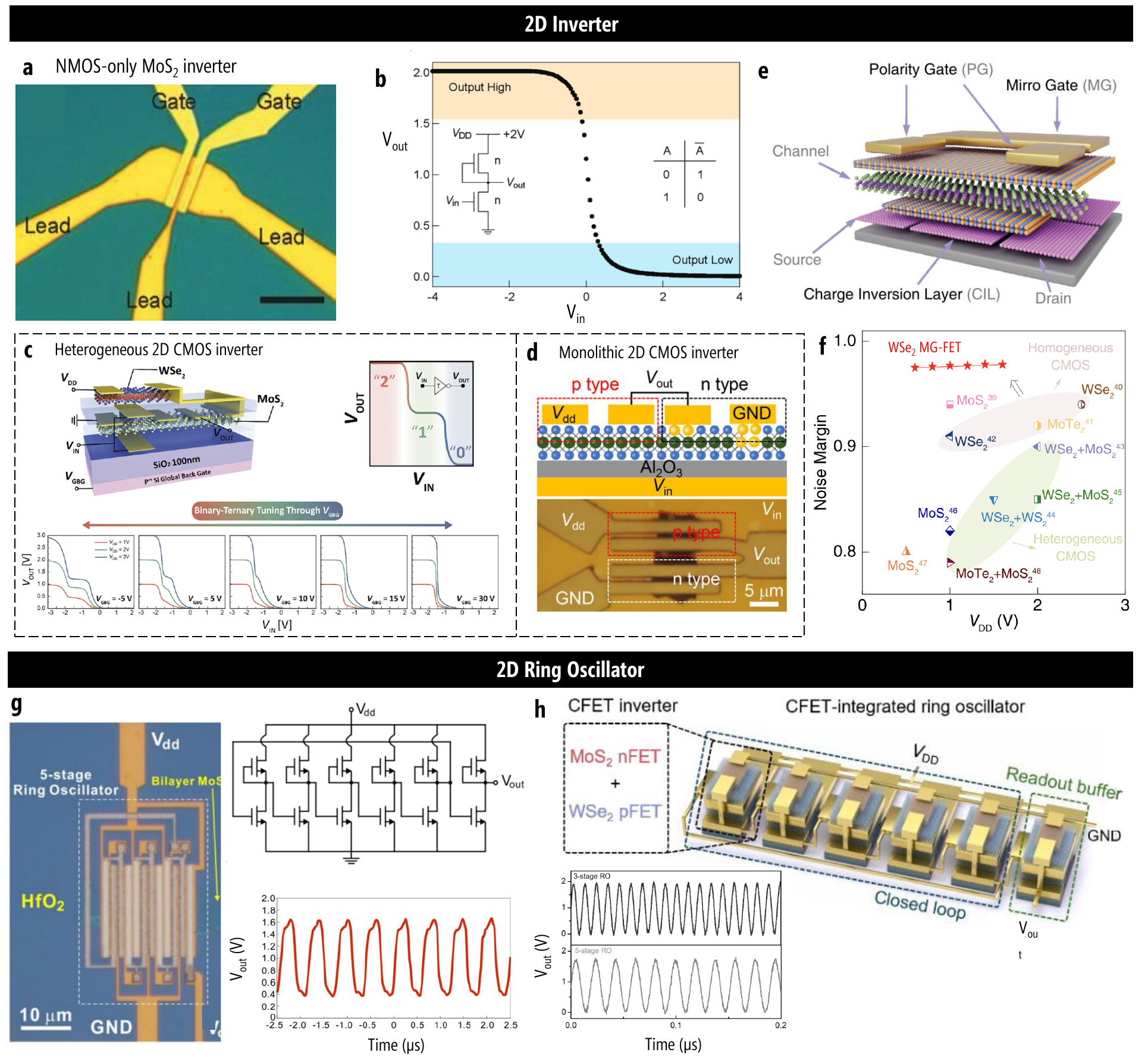}
    \caption{\textbf{2D logic gate cells.} 
    (a) 2D inverter microscopic image \cite{radisavljevicIntegratedCircuitsLogic2011}
    (b) 2D inverter schematic, VTC, and truth table \cite{radisavljevicIntegratedCircuitsLogic2011}
    (c) 3D model and characteristics of ternary CMOS (T-CMOS)\cite{leeElectricallyBinaryTernary2025}
    (d) Monolithic 2D CMOS inverter schematic and microscopic image \cite{liuDopingFreeMonolithic2D}
    (e) Mirror-gate FET (MGFET) 3D model \cite{yiCapacitiveAmplificationBoltzmannLimit}
    (f) MGFET noise margin comparison \cite{yiCapacitiveAmplificationBoltzmannLimit}
    (g) \ce{MoS2} 5-stage ring oscillator microscopic image, schematic, and oscillation plot \cite{wangIntegratedCircuitsBased2012}
    (h) \ce{MoS2}/\ce{WSe2} CFET inverter 3D model and oscillation plot \cite{kangHighkDielectricVan2025}
    }
    \label{fig:inverter_and_ro}
\end{figure*}

\subsubsection{2D-transistor-based inverters}

Various designs of inverters have been demonstrated using 2D transistors, covering both homogeneous and heterogeneous integration of primarily 2D TMDs. Earlier inverter demonstration uses NMOS-only \ce{MoS2} transistor operating in E-mode to achieve a maximal gain above 4 in this early prototype \cite{radisavljevicIntegratedCircuitsLogic2011} \figsub{fig:inverter_and_ro}{a, b}. The 2D inverter designed has been substantially improved in subsequent works. For instance, a 2D inverter with rail-to-rail operation, a mid-range threshold voltage, and a maximum gain of 344 has been introduced \cite{tianRailtoRailMoS2Inverters2022}. The inverter also exhibits a large noise margin of 98\% of $V_{dd}/2$ when $V_{dd}$ = 4 V. More recently, the gain of inverters implemented in a 2D RISC-V chip \cite{aoRISCV32bitMicroprocessor2025} reaches as high as 760 under $V_{dd}$ = 4 V operation with rail-to-rail output. Using DCFL approach composed of D-and E-mode transistors with different $V_\text{th}$ values, the inverters are designed to have a mid-range $V_\text{th}$ at 2 V.

Beyond NMOS logic families, 2D CMOS offer another route towards high-performance inverters. 2D CMOS inverter can be implemented using \emph{heterogeneous} approach which incorporates both \ce{MoS2} $n$-type transistors and \ce{WSe2} $p$-type transistors. Recent demonstration of heterogeneous 2D CMOS inverter under a vertical stacking architecture \cite{yiCapacitiveAmplificationBoltzmannLimit} \figsub{fig:inverter_and_ro}{c} suggest the potential of 2D transistors in complementary FET (CFET) design. Interestingly, the 2D CMOS inverter stacks can be dynamically switched via a global back gate to achieve ternary logic applications, thus suggesting novel logical functionality beyond Boolean algebra. Monolithic homogeneous 2D CMOS, which incorporates only one type of 2D semiconductor, has also been explored for inverter and circuit design due to their material simplicity. Recently, doping-free fabrication of $n$-type and $p$-type nature of \ce{WSe2} (or \ce{MoTe2}) transistors using a single type of contact metals but under different contact engineering, i.e. van der Waals versus direct deposition \cite{leeElectricallyBinaryTernary2025} has been demonstrated \figsub{fig:inverter_and_ro}{d}, thus concretely establishing the feasibility of high-quality inverter with high noise margin and high gain via monolithic 2D CMOS integration. 
Beyond the standard FET configuration, a mirror-gate \ce{WSe2} transistor \cite{yiCapacitiveAmplificationBoltzmannLimit} in which a charge inversion layer capacitive is generated via a polarity gate \figsub{fig:inverter_and_ro}{e} to amplify the gate control over the channel has been demonstrated. Such monolithic homogeneous device architecture enables 
sub-thermionic operation in a single device, and can achieve a high gain of 213 at 1 V in a monolithic homogeneous 2D CMOS inverter setup with record high noise margin as compared to other 2D CMOS inverters \figsub{fig:inverter_and_ro}{f}.
Overall, the demonstrations of broad variety of high-gain, high-noise-margin inverters provides a solid assurance on the compatibility of 2D semiconductor for complex circuit integrations.  

\subsubsection{Ring oscillator}

The performance of inverters is often further evaluated through ring oscillator (RO) measurements, which assess how well multiple inverters operate when connected in sequence. An RO consists of an odd number of inverters connected in a loop, causing the signal to propagate continuously through each stage. As the oscillation frequency is determined by the cumulative delay of all inverters, ROs serve as a sensitive benchmark of inverter speed, signal integrity, and stage-to-stage noise accumulation. A 5-stage RO oscillating at 0.52~MHz at $V_{DD}=1.15$V \cite{wangIntegratedCircuitsBased2012} based on \ce{MoS2} NMOS inverters represents a key early demonstration 2D-transistor-based RO \figsub{fig:inverter_and_ro}{g}. The operating frequency has since been significantly increased, reaching 13.12 MHz at $V_{DD}=15$V \cite{liLargescaleFlexibleTransparent2020} in a 5-stage RO, and 6.7MHz at $V_{DD}=5$~V \cite{tangLowPowerFlexible2023} in a 11-stage RO. Beyond NMOS-only inverters, 2D CMOS inverters have also been incorporated in RO design. For example, vertically-stacked $p$-type \ce{WSe2} and $n$-type \ce{MoS2} CMOS inverter have been employed to yield maximum frequencies of 91.5 MHz and 53 MHz for three-stage and five-stage ROs \figsub{fig:inverter_and_ro}{h}, respectively \cite{kangHighkDielectricVan2025}. This experiment further demonstrate the feasibility of 2D semiconductors in advanced architectures such as complementary FET (CFET) \cite{yoonEnablingAngstromEra2025}.

\begin{figure*}[t]
    \includegraphics[width=\textwidth]{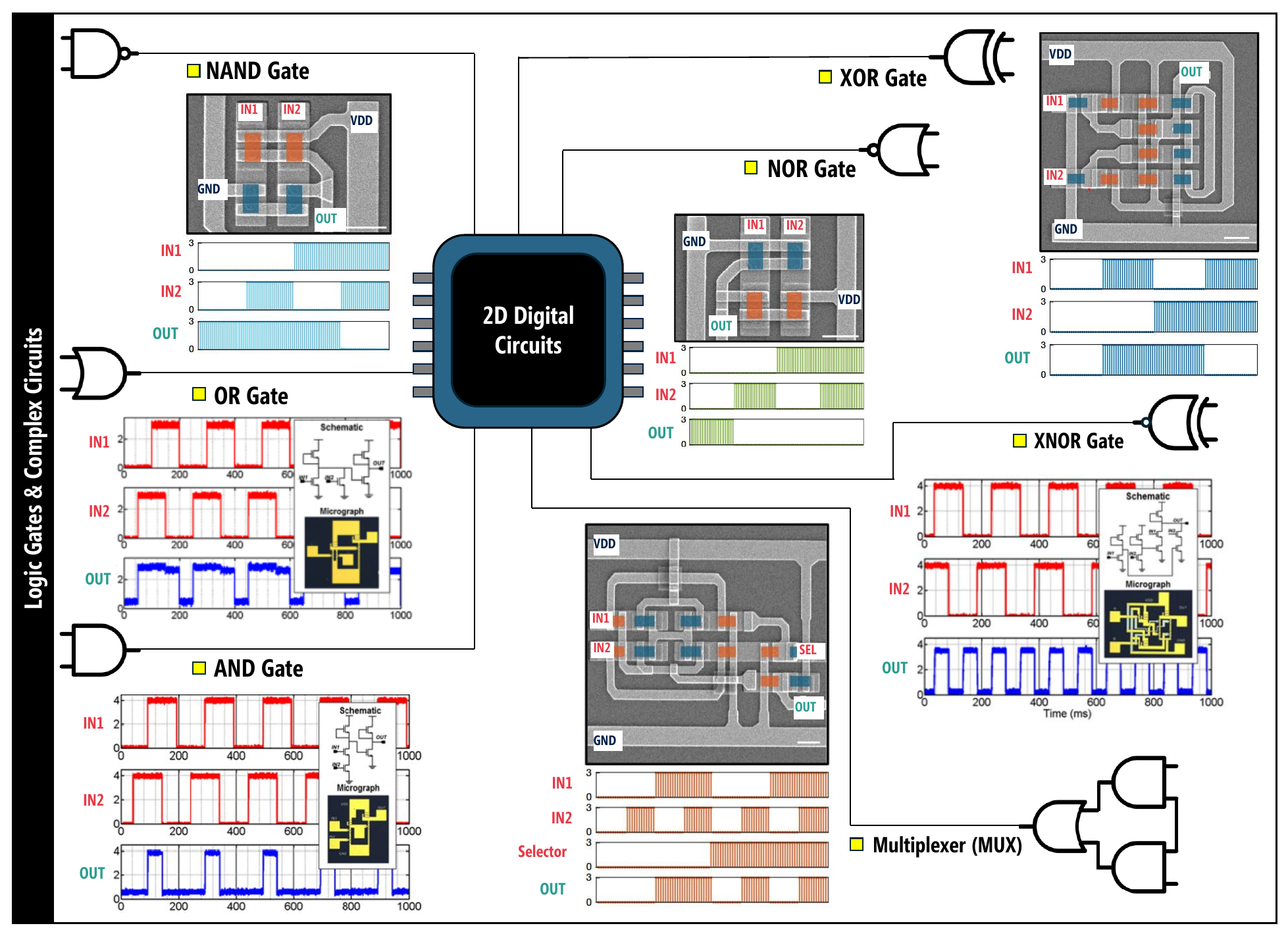}
    \caption{\textbf{2D logic gates and complex circuits} NAND gate, NOR gate, AND gate, OR gate,XOR gate, XNOR gate, and MUX cell \cite{yuDesignModelingFabrication2016, ghoshComplementaryTwodimensionalMaterialbased2025}
    }
    \label{fig:logiccell}
\end{figure*}

\subsubsection{Logic gates}

Logic gates are the core functional units of digital circuits, which perform boolean operations based on one or more input signals. Common gate types include NOT, NAND, NOR, AND, OR, XOR, and XNOR \fig{fig:logiccell}. These gates serve as the building blocks for constructing more complex circuit modules. For example, an AND gate outputs a logical high only when all inputs are high, enabling conditional operations in arithmetic and control logic. By combining these basic gates, large-scale 2D circuits and chips can be constructed.

Wide-range of 2D logic gates have been implemented with various logic families \cite{ yuDesignModelingFabrication2016, wachterMicroprocessorBasedTwodimensional2017, linSolutionprocessable2DSemiconductors2018, wangElectronicDevicesCircuits2019, liLargescaleFlexibleTransparent2020, tianRailtoRailMoS2Inverters2022, tangLowPowerFlexible2023, pengMediumscaleFlexibleIntegrated2024, aoRISCV32bitMicroprocessor2025, ghoshComplementaryTwodimensionalMaterialbased2025, wangMediumscaleIntegratedCircuits2026} \fig{fig:celltable}. NOR and NAND gates are typically focused on the early demonstration \cite{radisavljevicIntegratedCircuitsLogic2011, wangIntegratedCircuitsBased2012}, as they form a functionally complete set, i.e. any digital logic function can be synthesized from either NOR or NAND gates alone. These establishment illustrates the feasibility of multi-transistor logic in 2D devices and validate the underlying logic families: RTL \cite{radisavljevicIntegratedCircuitsLogic2011}, NMOS-only \cite{wachterMicroprocessorBasedTwodimensional2017, aoRISCV32bitMicroprocessor2025}, and CMOS \cite{ghoshComplementaryTwodimensionalMaterialbased2025}.

Beyond simple logic gates, more complex gates such as XOR, XNOR, and multiplexers (MUX) have been experimentally realized \cite{aoRISCV32bitMicroprocessor2025, ghoshComplementaryTwodimensionalMaterialbased2025}. These gates are essential for arithmetic operations, data routing, and signal selection-making them indispensable in the data path and control blocks of 2D digital systems.

It should be noted that certain combinational logic cells, including AOI and OAI gates, suffer from limited noise margins due to weak load devices in NMOS logic. Although gates such as AOI21 and OAI31 have been successfully demonstrated in the standard cell library of a 2D RISC-V microprocessor \cite{aoRISCV32bitMicroprocessor2025}, their constrained noise margins prevent the inclusion of more complex variants, such as AOI333, AOI222, and OAI221. This highlights the ongoing need to balance gate functionality with electrical robustness to achieve practical and high-performance 2D chips, with particular emphasis on improving noise margin.

\begin{figure*}[t]
    \includegraphics[scale=0.65]{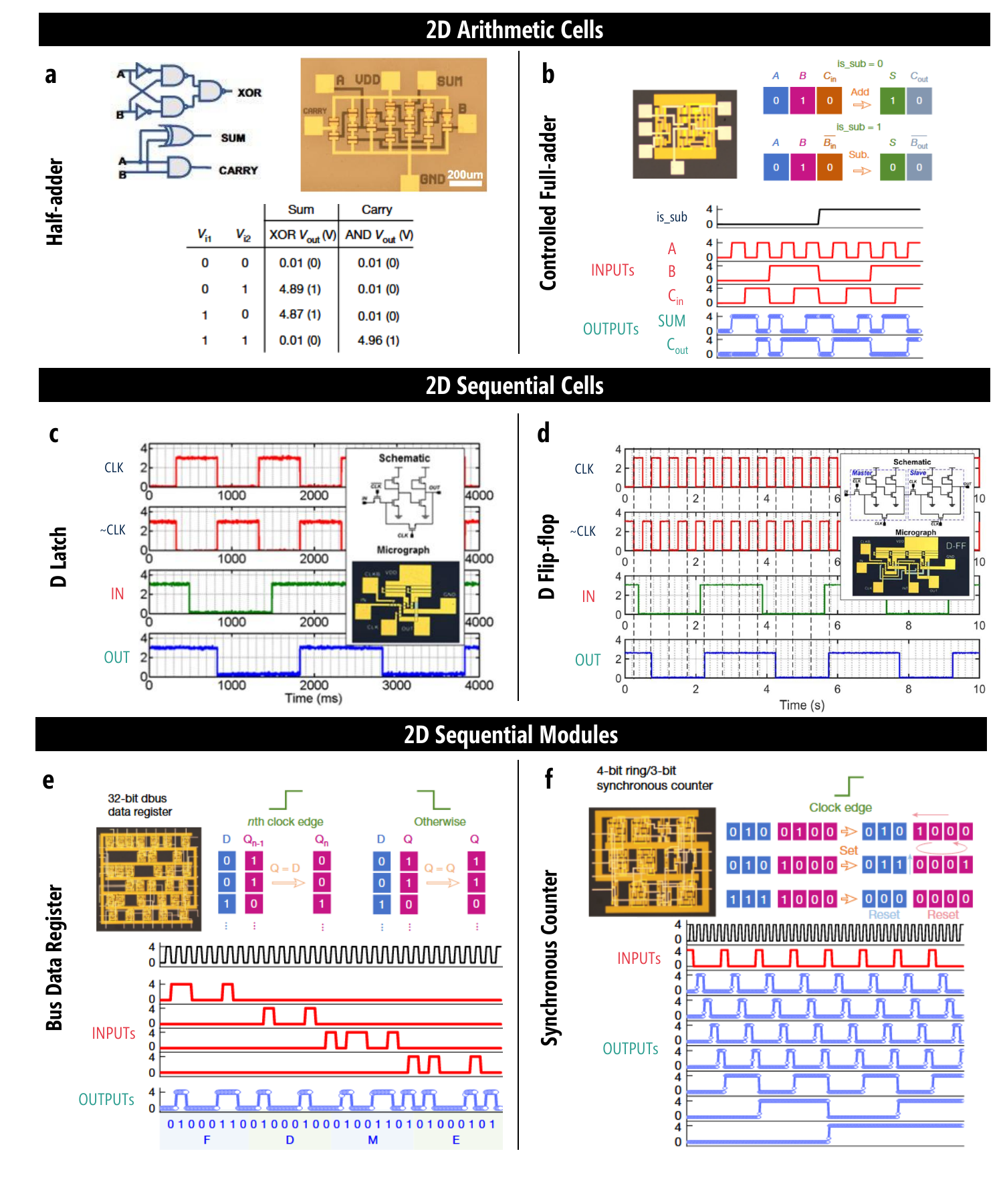}
    \caption{\textbf{2D computational building blocks: arithmetic cell as well as sequential cells and modules.} 
    (a) half-adder \cite{pengMediumscaleFlexibleIntegrated2024} 
    (b) full-adder \cite{aoRISCV32bitMicroprocessor2025}
    (c) D Latch \cite{yuDesignModelingFabrication2016}
    (d) D Flip-flop \cite{yuDesignModelingFabrication2016}
    (e) Bus Data Register \cite{aoRISCV32bitMicroprocessor2025} 
    (f) Synchronous Counter \cite{aoRISCV32bitMicroprocessor2025}
    }
    \label{fig:arithseq}
\end{figure*}

\subsubsection{Arithmetic cells}

Arithmetic cells perform the essential binary operations required in a processor’s arithmetic logic unit (ALU), including addition, subtraction, and basic signal combination. Common examples are the half-adder, full-adder, carry generator, and multiplier. These building blocks are indispensable for enabling computation, as they form the core datapath elements of microprocessor architectures.

Early demonstrations of 2D arithmetic cells focused on half-adders implemented using \ce{MoS2} transistors \cite{linSolutionprocessable2DSemiconductors2018}, which established the feasibility of performing arithmetic operations in flexible and large-area 2D integrated circuits \cite{pengMediumscaleFlexibleIntegrated2024} \figsub{fig:arithseq}{a}. More recently, microprocessor-scale 2D integration has advanced toward more functional designs. A notable example is the realization of a controlled full-adder, which extends a conventional full-adder by incorporating an additional input that selects between addition and subtraction operations \figsub{fig:arithseq}{b}. This controlled full-adder design enables both arithmetic functions within a single compact module, thus demonstrating that 2D logic can support multi-operation datapath components rather than isolated binary adders.

The successful realization of half-adders, full-adders, and multi-operation arithmetic units provides strong evidence that 2D semiconductors can be integrated into the higher-level datapath circuits required for full microprocessor design.

\subsubsection{Sequential cells}

Sequential cells are digital circuit elements capable of storing state information and are fundamental to implementing memory, control logic and finite-state machines. Unlike combinational gates whose outputs depend only on present inputs, sequential cells retain their previous output until updated. The simplest sequential cell is the SR latch, which holds its output until a signal is applied to either the set (S) or reset (R) input. An SR latch can be reconfigured into a D latch, which captures the input when the clock signal is high, or further into a D flip-flop (DFF), which captures data only at the clock edge. These components form the backbone of registers, counters, and state machines in modern processors.

Sequential cells have become a central focus in 2D standard-cell development \figsub{fig:arithseq}{c, d}. 2D D latches and DFFs have been experimentally demonstrated for test modules \cite{yuDesignModelingFabrication2016}, and their realizations in flexible electronics confirm that state-holding behavior can be preserved even under mechanical deformation \cite{pengMediumscaleFlexibleIntegrated2024}. This validates the feasibility of building reliable timing and control structures in both rigid and flexible 2D technologies.

Sequential cells are routinely incorporated into 2D microprocessor designs. In one microprocessor demonstration \cite{wachterMicroprocessorBasedTwodimensional2017}, D latches are employed as memory elements for storing intermediate results during instruction execution. In a more advanced integration effort, DFFs derived from a 2D standard-cell library were used to construct key system modules, including a 32-bit bus register \figsub{fig:arithseq}{e} and a 4-bit synchronous counter for the processor’s state machine \figsub{fig:arithseq}{f} \cite{aoRISCV32bitMicroprocessor2025}. Additionally, DFFs have been incorporated as buffer elements within processing pipelines of CMOS-compatible 2D logic systems \cite{ghoshComplementaryTwodimensionalMaterialbased2025}, thus enabling stable data transfer between stages and demonstrating that 2D sequential logic can support coordinated and clocked operations across multistage circuits. 2D semiconductor circuits are thus capable not only of implementing basic memory elements but also of enabling the full range of sequential logic required for scalable computing architectures.

\begin{figure*}[t]
    \includegraphics[scale=0.47]{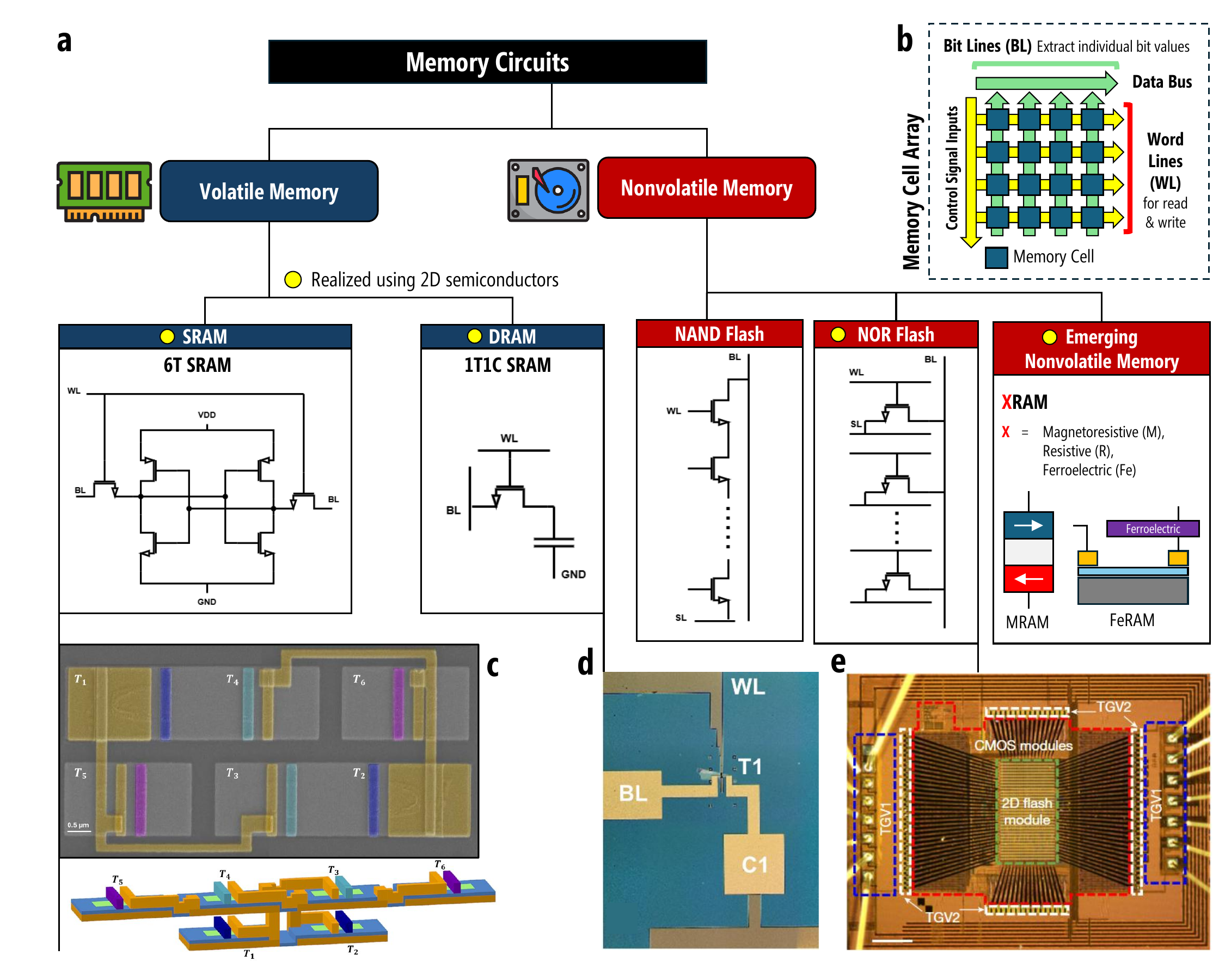}
    \caption{\textbf{2D Memory Circuits} (a) Memory circuit hierachy and cell schematic: SRAM, DRAM, NAND flash, and NOR flash. WL: word line, BL: bit line, SL: source line (b) Memory cell with the description of bit lines and word lines (c) planar implementation of 2D 6T-SRAM \cite{sadafEnablingStaticRandomaccess2025}, (d) 3D implementation of 2D 6T-SRAM \cite{sadafEnablingStaticRandomaccess2025}, (e) 2D 1T1C cell implementation \cite{kshirsagarDynamicMemoryCells2016}, (f) 2D NOR flash implementation \cite{liuFullfeatured2DFlash2025}}
    \label{fig:memory}
\end{figure*}

\subsection{Memory Circuits}

Memory circuits play a critical role not only in digitial computing but also pave the cornerstone for artificial intelligence \cite{woltersMemoryAllYou2024}. The integration of 2D semiconductors into memory chips has been intensively studied recently \cite{liuFullfeatured2DFlash2025, qinTwoDimensionalMaterialsUltimate2025, xiangSubnanosecondFlashMemory2025}. For completeness, we briefly introduce the major classes of 2D-semiconductor-based memory circuits in this subsection. Interested readers are suggested to consult comprehensive reviews such as Ref.\cite{maCircuitLevelMemoryTechnologies2022}.

Generally, memory circuits store binary information in either volatile or non-volatile forms \cite{harasztiIntroductionCMOSMemories2002, maCircuitLevelMemoryTechnologies2022} \figsub{fig:memory}{a}. Volatile memories lose data once power is removed, whereas non-volatile memories retain information without continuous power. Among volatile memories, static random-access memory (SRAM) is the dominant technology due to its fast access speed and low latency. A typical SRAM cell uses a 6-transistor (6T) architecture, which is consisting of two cross-coupled inverters forming a bistable latch and two access transistors controlling reads and writes. Because the stored bit is maintained through the positive feedback of the inverter pair rather than stored charge, SRAM is inherently volatile and optimized for high-speed cache and register applications. The design also underscores the need for high-noise-margin 2D inverters to ensure robust 2D SRAM operation. 

Dynamic random-access memory (DRAM) is another memory circuit type for higher memory density than SRAM. The DRAM use capacitor to store the charge instead. Therefore, the DRAM memory cell such as 1-transistor-1-capacitor (1T1C) \cite{maCircuitLevelMemoryTechnologies2022} is smaller than SRAM counterpart but prone to charge leakage. Thus, the cell require refreshing to store the data consistently. Short-channel and thin 2D devices have a potential to maximize memory density of DRAM development.

In contrast, flash memory serves as a representative non-volatile technology. Flash devices store data by trapping charge in a floating gate or charge-trap layer, shifting the transistor threshold voltage to encode `0' or `1'. Such mechanism allows data retention for years without power but results in slower write operations and limited endurance as compared to SRAM. Flash memory arrays are commonly organized as NOR flash or NAND flash \cite{micheloniNANDFlashMemories2010}, where the naming reflects the logical structure used for word-line and bit-line connectivity. Because of the memory storage capacity requirement, strategy to achieve dense integration of high-quality NOR and NAND are crucial for enabling 2D semiconductor flash memory technologies.

To integrate a large memory circuit, the memory cells are arranged in the form of an array at the intersections of word lines (WLs) and bit lines (BLs) \figsub{fig:memory}{b}. A WL is a circuit line that carries control signals to enable read and write operations. In contrast, a BL is a circuit line that connects to each individual bit of a word in memory. By combining these intersecting lines on each memory cell, the circuit enables both control signaling and bit-level data access at every cell. This architectural design also serves as a fundamental concept for 2D memory circuit design.

Recent work demonstrates significant progress in 2D volatile memory. A monolithic 3D SRAM block consisting of 6,144 transistors has been fabricated using 2D semiconductor layers \figsub{fig:memory}{c}, achieving kilobit-scale storage \cite{sadafEnablingStaticRandomaccess2025}. Multi-tier stacking strategies have been employed to reduce cell footprint: 2-tier and 3-tier SRAM configurations achieve 40\% and 70\% reductions in cell area, respectively, compared to a single-tier design. It should be noted that although these architectural refinements improve density, they also introduce fabrication challenges. For example, yield decreases slightly from 98\% in the first tier to 97\% in the second tier. Nevertheless, these pioneering works demonstrate that 2D semiconductors can support high-density vertically integrated volatile memory. In the case of DRAM, \ce{MoS2} DRAM has been fabricated using 1T1C and 2-transistor (2T) architectures \cite{kshirsagarDynamicMemoryCells2016}. The retention times before the cells leak charge are 0.25 and 1.3 s for 1T1C \figsub{fig:memory}{d} and 2T, respectively, with a leakage current of 1–2 \si{fA/\micro\meter}. This study shows the potential of 2D materials to fabricate DRAM circuits.

For non-volatile memory circuit, a fully featured 2D NOR flash chip has been integrated onto a commercial CMOS substrate \cite{liuFullfeatured2DFlash2025} \figsub{fig:memory}{e}. This work introduces an atomic-device-to-chip framework, ATOM2CHIP, which bridges atomic-scale \ce{MoS2} device engineering with system-level chip integration \figsub{fig:memory}{e}. Importantly, standard chip-verification procedures are implemented, thus allowing the compatibility between 2D device fabrication and CMOS process. The resulting chip demonstrates a 20 ns operation speed, 0.644 pJ per-bit energy consumption, and a projected 10-year lifetime at 54.8 $^{\circ}$C, achieving a fabrication yield of 94.34\%, thus revealing the potential of 2D semiconductors in non-volatile memory chips.

\begin{figure*}[t]
    \includegraphics[scale=0.55]{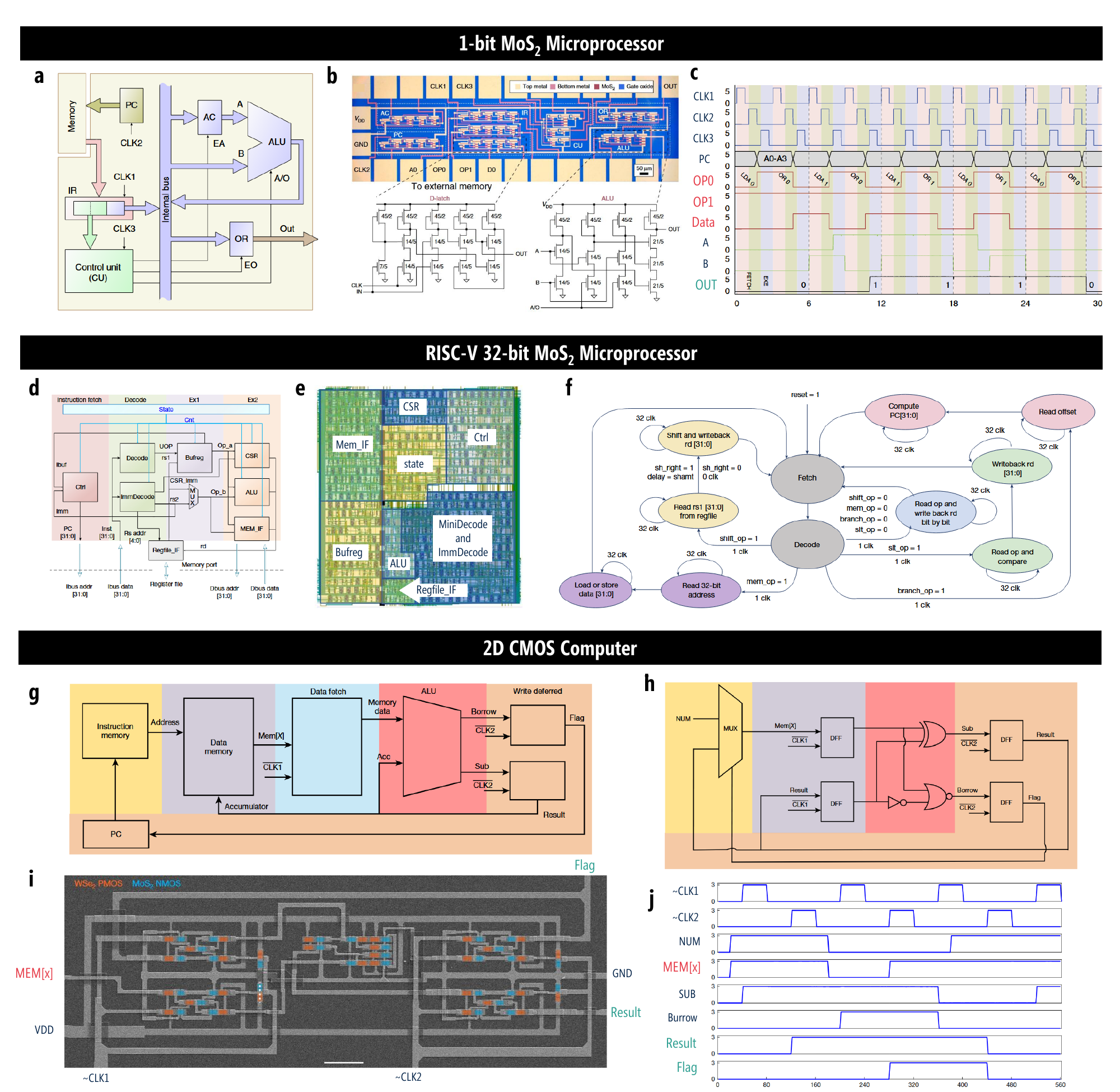}
    \caption{\textbf{2D Mircroprocessors with schematic, computing process and microscopic image} 
    (a, b, c) 1-bit \ce{MoS2} microprocessor \cite{wachterMicroprocessorBasedTwodimensional2017}
    (a) architecture schematic 
    (b) microscopic image and its circuit schematic 
    (c) timing diagram 
    (d, e, f) RISC-V 32-bit \ce{MoS2} microprocessor \cite{aoRISCV32bitMicroprocessor2025}
    (d) architecture schematic 
    (e) microscopic image with module separation highlight
    (f) state machine
    (g, h, i, j) 2D CMOS \ce{MoS2}/\ce{WSe2} Computer \cite{ghoshComplementaryTwodimensionalMaterialbased2025}
    (g) architecture schematic for OISC instruction processing
    (h) architecture schematic for RSSB instruction processing
    (i) SEM image of OISC
    (j) timing diagram for RSSB operation
    }
    \label{fig:microprocessor}
\end{figure*}

\subsection{2D semiconductor microprocessors}

Driven by the successful integration of 2D semiconductors into various logic gates, arithmetic units, sequential elements and memory circuits, the next major challenge is the realization of microprocessor chips based on 2D semiconductors. Recent breakthroughs including 1-bit microprocessors \cite{wachterMicroprocessorBasedTwodimensional2017, guoTowardsfoundryStrategyCreating2026}, one-instruction-set computers (OISC) \cite{ghoshComplementaryTwodimensionalMaterialbased2025} and 32-bit RISC-V architectures \cite{aoRISCV32bitMicroprocessor2025} demonstrate that 2D semiconductors can support not only circuit-level functionality but also full system-level computing. These demonstrations provide strong assurance that 2D semiconductors are viable candidates for future computing. In this section, we review the progression of 2D microprocessor development and highlight the key architectural strategies, device innovations, and fabrication advances that have enabled these early prototypes of 2D chips.

\subsubsection{1-bit microprocessors with \texorpdfstring{\ce{MoS2}}{MoS2} transistors}

The demonstration of a 1-bit \ce{MoS2}-based microprocessor in 2017 \cite{wachterMicroprocessorBasedTwodimensional2017} represents a landmark advancement towards 2D chips. The 1-bit microprocessor employs inverters and NAND gates engineered for rail-to-rail operation through asymmetric sizing of load and drive transistors. These standard cells are composed into the core modules of a rudimentary microprocessor, including ALU program counter (PC), instruction register (IR), and control unit (CU) \figsub{fig:microprocessor}{a, b}. Using 115 transistors and three clock signals, the system has the capability to execute four three-bit opcode instructions, i.e., NOP, AND, OR, and LDA \figsub{fig:microprocessor}{c}. Although minimalist design, the architecture is inherently scalable where a multi-bit processor can be formed by arranging multiple 1-bit slices in parallel.

This pioneering also reveals the key bottlenecks in 2D microprocessor engineering. While individual modules were fabricated with yields approaching 80\%, integration into a full system reduced overall yield to only a few percent. The primary source of defects was attributed to imperfections introduced during \ce{MoS2} film transfer \cite{wachterMicroprocessorBasedTwodimensional2017, shanmugamReviewSynthesisProperties2022}, suggesting that direct-growth fabrication could substantially improve reliability. A second limitation is the operating speed, which is constrained by the pull-up transistor characteristics. The maximum clock frequency is estimated to be 2-20 kHz, thus necessitating improvements in mobility and drive current. For instance, DCFL logic families can solve this issue without large channel width.

\subsubsection{RISC-V microprocessor with \texorpdfstring{\ce{MoS2}}{MoS2} transistors}

One of the most advanced (and complex) 2D microprocessors to date is the RV32-WUJI \ce{MoS2}, a RISC-V 32-bit microprocessor. The design of this 2D chip is enabled by a self-developed standard cell library comprising 25 logic units, ranging from basic inverters to complex sequential cells such as D flip-flops. Notably, the library includes gate types such as OAI, AOI, and NAND3B, which are rarely used in earlier experimental chips but are essential for constructing compact and scalable industrial-grade logic. The inclusion of these advanced cells significantly enhances logic density and circuit expressiveness, marking a major step toward manufacturable 2D system-on-chip designs.

The RV32-WUJI processor integrates approximately 5,900 transistors to realize a full set of RISC-V modules, including an ALU, state register, and control module \figsub{fig:microprocessor}{d, e}. Operationally, the processor can be viewed as a large state machine \figsub{fig:microprocessor}{f}, which coordinates sequential execution of instructions on a bit-serial datapath.

Despite its complexity relative to earlier 2D processors, the design still exhibits several limitations compared to modern silicon RISC-V architectures. First, the processor adopts a purely sequential, bit-serial architecture, thus processing only one bit per clock cycle; consequently, A full 32-bit operation requires 32 cycles. Second, the processor lacks architectural enhancements such as pipelining, caches, branch prediction and interrupt handling—all features that substantially boost throughput in commercial microprocessors. Finally, the architecture reflects the early stage of 2D integration, where fabrication yield, device variability, and interconnect parasitics still constrain system-level performance. Nevertheless, RV32-WUJI represents a clear milestone that demonstrates the feasibility of large-scale instruction-set-compatible processor chips using 2D semiconductors.

\subsubsection{2D CMOS computing chip}

In contrast to RV32-WUJI 2D chips, CMOS-based 2D microprocessor has been recently constructed by the complementary integration of \ce{WSe2} PMOS and \ce{MoS2} NMOS transistors \cite{ghoshComplementaryTwodimensionalMaterialbased2025}. This CMOS computing chip comprises approximately 1,000 PMOS and 1,000 NMOS devices, and is capable of implementing a one-instruction-set computer (OISC) based on the reverse subtract and skip if borrow (RSSB) instruction. The architecture incorporates dedicated hardware for RSSB execution \figsub{fig:microprocessor}{h}, as well as data management, control and computation modules \figsub{fig:microprocessor}{g, i}. Instruction execution is synchronized using two clock signals \figsub{fig:microprocessor}{j}, thus enabling deterministic sequencing of operations.

The demonstration of 2D CMOS chip highlights several advantages of CMOS logic for 2D integrated circuits. The complementary design achieves extremely low static and dynamic power consumption, at the picowatt and picojoule levels, respectively. Its transistor-efficient layouts provide higher area efficiency than 2D processors based on NMOS-only logic families. Furthermore, identical device geometries can be used across PMOS and NMOS transistors, enabling high noise margins and more standardized cell design—features that closely parallel mature CMOS design practices in silicon.

Although this CMOS 2D chip prototype supports only a single instruction, its reliable CMOS operation, modular structure and clean logic design indicate strong potential for scaling toward multi-instruction and multi-bit architectures. As such, this CMOS 2D chip prototype serves as an important demonstration that 2D semiconductors can support not only logic compatibility but also full CMOS functionality at the system level.

\begin{figure*}[t]
    \includegraphics[scale=0.53]{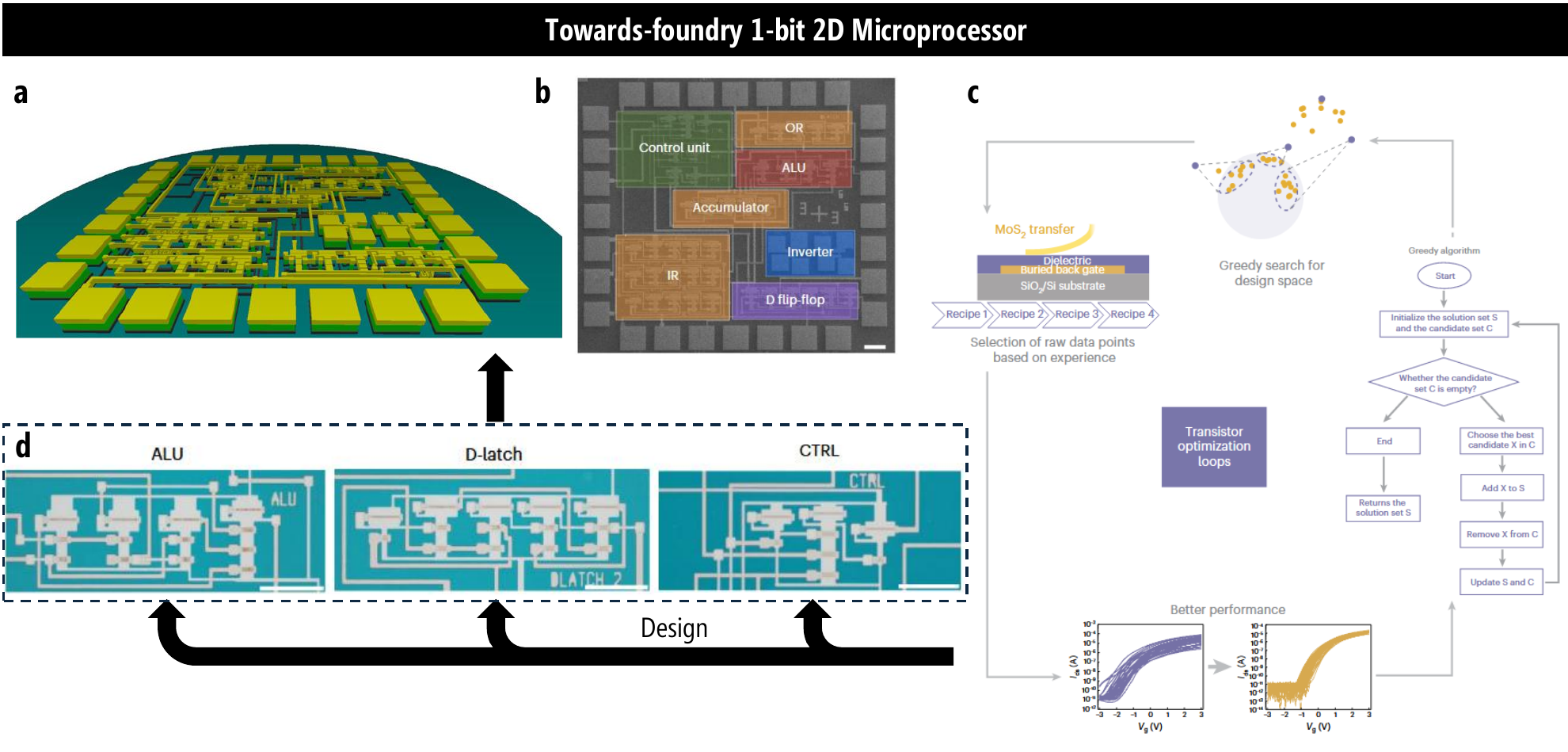}
    \caption{\textbf{Towards-foundry 2D microprocessor} \cite{guoTowardsfoundryStrategyCreating2026} (a) 3D model of 1-bit 2D microprocessor (b) microscopic image (c) transistor optimazation loop flowchart contributing with greedy algorithm, called "towards-foundry" strategy (d) Circuit implementation microscopic images: ALU module, D-latch, and CTRL module.
    }
    \label{fig:towards}
\end{figure*}

\subsubsection{Towards-foundry 2D microprocessor}

Recently, a significant milestone in large-scale 2D chip demonstration was achieved through a "towards-foundry" manufacturing strategy, culminating in the realization of a fully interconnected 1-bit microprocessor \figsub{fig:towards}{a, b} based on transferred monolayer MoS${2}$ \cite{guoTowardsfoundryStrategyCreating2026}. Diverging from traditional small-batch laboratory fabrication, this approach utilized an iterative, multidimensional optimization loop guided by a greedy algorithm \figsub{fig:towards}{c}, spanning material growth, layout design, transfer, device fabrication, and chip probing across approximately 130 batches. This rigorous process enabled the epitaxial growth of highly uniform MoS${2}$ films scaled up to four-inch wafers, yielding single transistors with nearly 100\% reliability, on/off ratios exceeding 10$^{7}$, and inverter voltage gains averaging 400. Building on these robust discrete devices, researchers achieved exceptionally high fabrication yields for essential circuit modules \figsub{fig:towards}{d}, including arithmetic logic units (ALUs) at 96.5\%, control units at 79.5\%, and D-latches at 61.5\%. Ultimately, these modules were seamlessly integrated into a fully functional, 120-transistor microprocessor comprising an instruction register, output register, accumulator, ALU, control unit, and D flip-flops. Capable of executing cyclic instruction sequences with excellent signal integrity and full-swing voltage logic, this 2D MoS$_{2}$ CPU demonstrates a lower normalized power consumption compared to early-generation silicon microprocessors (such as the Intel 4004), forcefully validating the scalability and energy-efficient potential of 2D semiconductors for next-generation computing architectures.

\begin{figure*}[t]
    \includegraphics[scale=0.53]{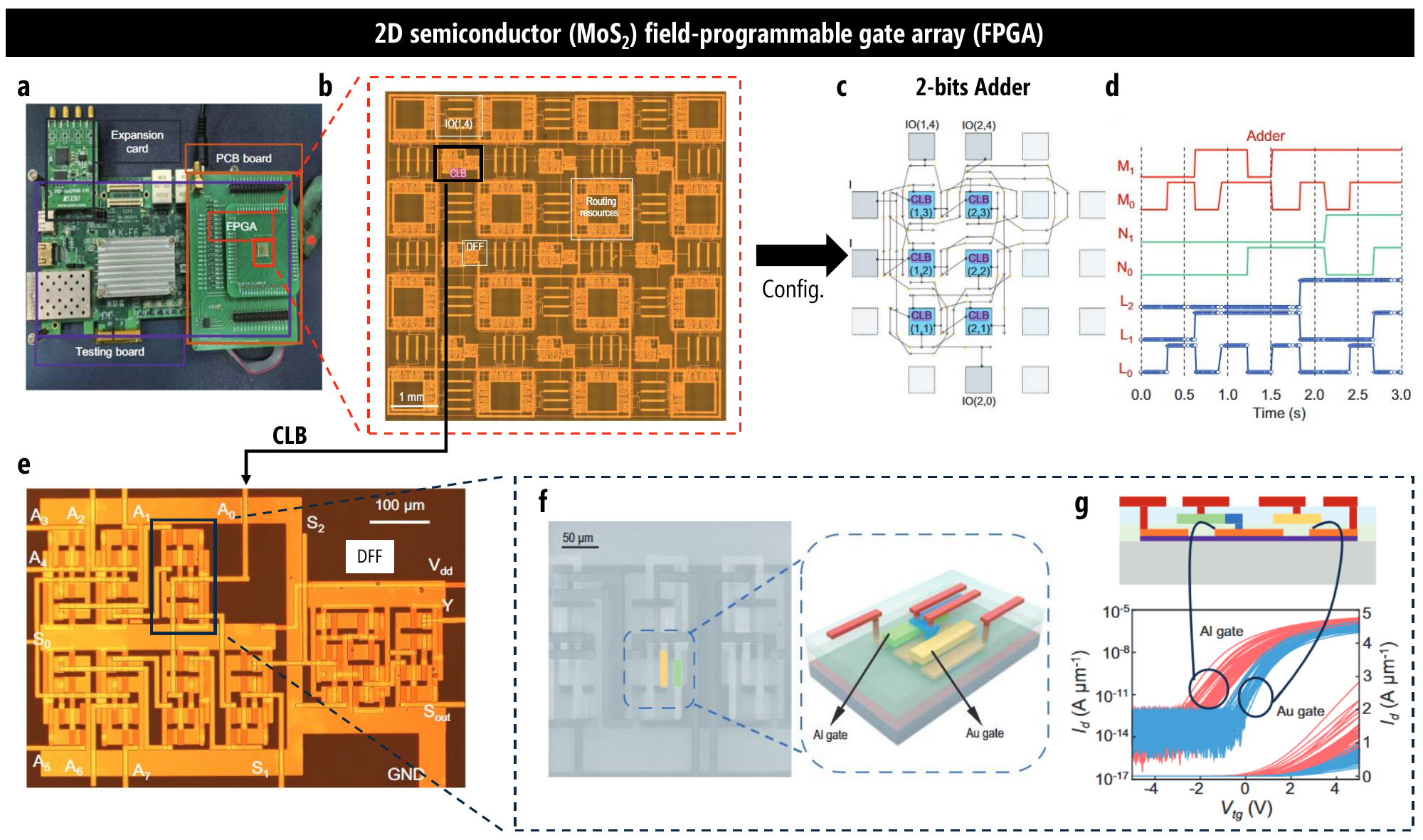}
    \caption{\textbf{2D semiconductor FPGA} \cite{sunFieldprogrammableGateArray2025} (a) Testing platform (b) microscopic image with the highlights of components: configurable logic block (CLB), input-output (IO), and routing resources (c) 2-bit adder use case example schematic (d) the digital signal inputs and outputs of configured FPGA 2-bit adder (e) CLB microscopic image (f) SEM image and 3D schematic of DCFL logic block (g) crossection image of the DCFL logic block with the transfer characteristic curve of FETs
    }
    \label{fig:fpga}
\end{figure*}

\subsection{2D semiconductor field-programmable gate array (FPGA)}

Field-programmable gate array (FPGA) represents a versatile class of programmable logic that offers significant scalability and flexibility compared to traditional application-specific integrated circuits (ASICs). Unlike ASICs, FPGAs can be configured by users post-fabrication using hardware description languages to implement specific logic functions, making them essential for rapid prototyping and customized hardware solutions \cite{sunFieldprogrammableGateArray2025}. Their inherent parallelism and reconfigurability have led to widespread adoption in fields such as digital signal processing, network communication, and deep learning acceleration. Furthermore, the modular nature of FPGAs makes them an ideal platform for integrating emerging 2D semiconductor technologies into complex and large-scale functional circuits.

Building on these advantages, researchers have recently demonstrated the first FPGA constructed from wafer-scale MoS$_2$, which integrates approximately 4,000 2D FETs \cite{sunFieldprogrammableGateArray2025} \figsub{fig:fpga}{a, b}. This 2D FPGA utilizes a silicon-compatible top-gate process and features an architecture comprising nine configurable logic blocks (CLBs), routing resources, and a dedicated memory array \figsub{fig:fpga}{b} where each cell adopts a compact 2T0C DRAM structure. The platform's reconfigurability was successfully validated through the implementation of functional circuits, including adders \figsub{fig:fpga}{c, d}, multipliers, and counters. Remarkably, the 2D FPGA exhibit exceptional irradiation resistance up to 10 Mrad, which is significantly higher than silicon CMOS \cite{leeDesignHighReliability2020}. In CLB \figsub{fig:fpga}{e}, the current 2D FPGA unit design \figsub{fig:fpga}{f} is based on DCFL family using only MoS$_2$ with different gate metals: Al for D-mode NMOS and AU for E-mode NMOS \figsub{fig:fpga}{g}. The CMOS integration of 2D FPGA designs remains an open challenge, and achieving it could pave the way for significantly more compact and high-performance 2D integrated systems.

\subsection{2D semiconductors in emerging unconventional computing}

Beyond general-purpose digital processors, 2D semiconductors have also enabled a growing class of application-specific and unconventional computing architectures \cite{lemme2DMaterialsFuture2022, huang2DSemiconductorsSpecific2022}. These systems are typically designed for specialized tasks or exploit device physics that extend beyond conventional CMOS operation, such as brain-like computing \cite{kim2DMaterialsbased3D2024}, machine vision \cite{mennelUltrafastMachineVision2020} and compute-in-memory \cite{quheAsymmetricConductingRoute2024, wangInmemoryComputingArchitecture2021}. In this section, we briefly review such computing system and we recommend interested readers to consult specialized reviews on this rapidly evolving computing paradigm \cite{liuTwodimensionalMaterialsNextgeneration2020, dasTransistorsBasedTwodimensional2021}.

A significant thrust in this domain is neuromorphic computing, where 2D devices emulate neural or synaptic behaviors. Demonstrations include neuristor-based logic elements \cite{chenLogicGatesBased2021} and memtransistor arrays \cite{doddaAllinoneBioinspiredLowpower2022}, both of which leverage the inherent switching dynamics of 2D materials to mimic biological computation. In parallel, logic-in-memory concepts have been realized using field-gate FETs (FGFETs) \cite{bertolazziNonvolatileMemoryCells2013, migliatomaregaLogicinmemoryBasedAtomically2020, zouTwoDimensionalTunnelingMemtransistor2024}, as well as more recent dual logic-in-memory architectures \cite{niuDuallogicinmemoryImplementationOrthogonal2024}, which aim to reduce data-transfer bottlenecks by collocating computation and storage. 2D materials have also found application in in-sensor and edge computing. Image-processing arrays based on 2D optoelectronic devices \cite{zengApplicationspecificImageProcessing2022} and curved neuromorphic image sensors \cite{choiCurvedNeuromorphicImage2020} illustrate how 2D semiconductors can enable compact, low-power computational vision systems. Additionally, demonstrations of 2D perceptrons for hardware-accelerated artificial intelligence \cite{migliatomaregaLowPowerArtificialNeural2022} highlight the potential of 2D devices in machine-learning inference and lightweight AI accelerators.

These emerging computing systems showcase the diversity of application-specific computing paradigms enabled by 2D semiconductors. The low-power operation and compatibility with heterogeneous integration make 2D semiconductors promising for future edge, neuromorphic and logic-in-memory computing.

\begin{figure*}[t]
    \includegraphics[scale=0.38]{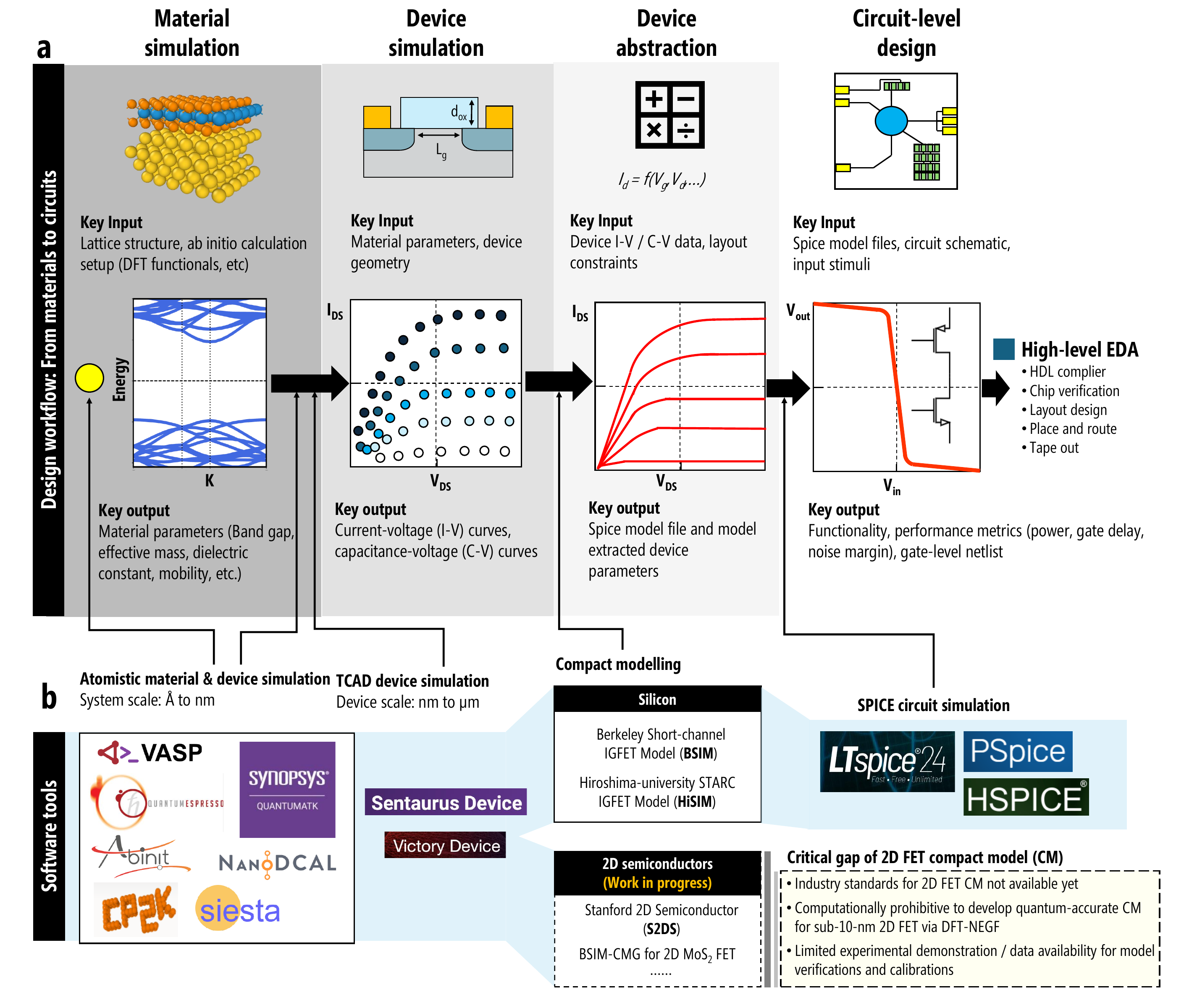}
    \caption{\textbf{Chip Design Flow from material-level to circuit-level} (a) Key inputs, outputs and processes on each design level (b) software and tools for doing simulation and modeling.}
    \label{fig:chipdesignflow}
\end{figure*}

\section{Compact Modeling for 2D Circuits and Chips}

The design of 2D semiconductor chips relies on a hierarchy of modeling and simulation tools that bridge atomic-level material physics to system-level electronic design automation (EDA). This multiscale workflow is essential because no single simulation method spans all relevant length scales, from Ångström-scale interfaces to fully integrated microprocessors. In this section, we first review the `atomistic-to-chip' design ecosystem, encompassing DFT, quantum transport, device simulation, compact modeling, and circuit/system simulation, followed by a detailed discussion on compact model development for 2D transistors and its importance for chip-scale integration.

\subsection{Chip Design Flow: A Primer for System Integrations}

Modern semiconductor design relies on a hierarchical modeling ecosystem that connects atomic-scale physics to full chip implementation. This hierarchy is particularly important for 2D semiconductors, whose atomic thickness, strong quantum confinement, and interface sensitivity demand simulation methods beyond those traditionally used for silicon. At the material level, density functional theory (DFT) and related first-principles techniques are used to extract fundamental quantities such as band structure, effective mass, defect energetics, electrical mobility and interface physics \cite{wangHighThroughputComputationalScreening2022, carvalhoComputationalMethods2D2021, ponceFirstprinciplesCalculationsCharge2020}. These atomistic insights are particularly useful in material optimization and selection, which pave the foundation for the subsequent design workflow \cite{haHighthroughputScreening2D2024}. Charge carrier transport models form the core of the device-level design workflow. For long-channel device, semiclassical TCAD tools, typically solving solving drift-diffusion (DD) or hydrodynamic equations, can capture various important effects such as electrostatics, mobility degradation, self-heating and parasitic resistances \cite{marinModelingElectronDevices2018}. Derived parameters from atomistic \textit{ab initio} methods (bandgap, effective mass, interface dipoles) can be incorporated in TCAD simulators for better capturing the physical properties of 2D semiconductors. As devices shrink toward the angstrom-scale technology nodes with near-or sub-10-nm physical gate lengths, classical carrier transport models no longer hold. Quantum transport simulations based on DFT-NEGF formalism are needed to capture quantum mechanical effects including tunneling that dominate ultrascaled 2D transistors \cite{quheSub10NmTwodimensional2021}. 

However, TCAD and DFT-NEGF are not suitable for full-circuit simulations due to their tremendous computational cost. To scale from device-level behavior to functional circuits, the extracted device characteristics must be abstracted into \textit{compact models}. Compact models express transistor operation through analytical or semi-empirical equations that are implemented in languages such as Verilog-A for SPICE-based simulation. The compact models therefore serve as the critical bridge between device physics and circuit design, enabling fast evaluation of logic gates, arithmetic blocks, memory cells and sequential elements using workflows that are familiar to CMOS circuit designers.

Once compact models are available, standard cells can be designed using SPICE-based circuit simulations, which are subsequently used for gate-level netlist generation and block-level verification. These simulations allow the assessment of delay, power, noise margins and signal integrity across large circuit. At the highest abstraction level, compact models are fed into system-level EDA workflows involving register-transfer level (RTL) coding in hardware description language (HDL) including Verilog or VHDL, logic synthesis, placement and routing, timing analysis, and physical verification. Such a multiscale design workflow, covering from atomistic simulation to device modeling, these design flows form an essential platform for integrating 2D semiconductor into future computing technologies.

\subsection{Why compact modeling is essential for 2D semiconductor circuits?}

Transistors are the key building blocks of computing chip. The physics, design and modeling of transistors thus lie at the heart of modern semiconductor design tools. For 2D semiconductors whose behavior is governed by atomically thin channels, strong quantum confinement, and interface-dominated electrostatics, significant efforts have been focused on the atomistic simulations and TCAD-level device modeling to capture their intrinsic material and transport characteristics \cite{silvestriHierarchicalModelingTCAD2023}. TCAD simulators, while computationally faster, rely on drift-diffusion or hydrodynamic models and empirical mobility or doping formulations developed for bulk or ultra-thin-body silicon. These assumptions break down for 2D devices, where transport is strongly affected by 2D density-of-states, contact-limited injection, monolayer electrostatics, and nanoscale tunneling \cite{knoblochModeling2DMaterialBased2023}. Directly applying silicon-based TCAD frameworks could thus lead to inaccurate predictions of current, capacitance, threshold voltage, switching delay and noise behavior in 2D transistors.

To translate device physics into functional circuits, \emph{compact models} are indispensable \cite{knoblochModeling2DMaterialBased2023}. Compact models provide analytical or semi-empirical equations that reproduce the key current–voltage and charge-voltage characteristics of a device while remaining computationally efficient for circuit-level simulations. TCAD simulators, while computationally faster, rely on drift-diffusion or hydrodynamic models and empirical mobility or doping formulations developed for bulk or ultra-thin-body silicon. These assumptions break down for 2D devices, where transport is strongly affected by 2D density-of-states, contact-limited injection, monolayer electrostatics, and nanoscale tunneling \cite{knoblochModeling2DMaterialBased2023, wangRoad2DSemiconductors2022, shenTrend2DTransistors2022}. Directly applying silicon-based TCAD frameworks could thus lead to inaccurate predictions in 2D transistors. Atomistic methods such as DFT-based NEGF can provide detailed quantum-mechanical insights, but they are computationally prohibitive for simulating logic gates or complex circuits which contains large number of interconnected transistors. 

The lack of industrial-grade compact models is thus a critical bottleneck of 2D semiconductors. Accurate compact models for 2D semiconductors is essential for integrating 2D transistors into the established EDA workflows, which is especially critical in performing timing analysis, noise margin evaluation, or power-delay optimization using SPICE environments \cite{tumaCircuitSimulationSPICE2009}. Importantly, compact models allow the computational benchmarking against silicon, as well as the optimal design of large-scale circuits. Compact modeling is thus a foundational enabler for advancing 2D semiconductors from device-level demonstrations to scalable chip-level integration.

\begin{figure*}[t]
    \includegraphics[scale=0.55]{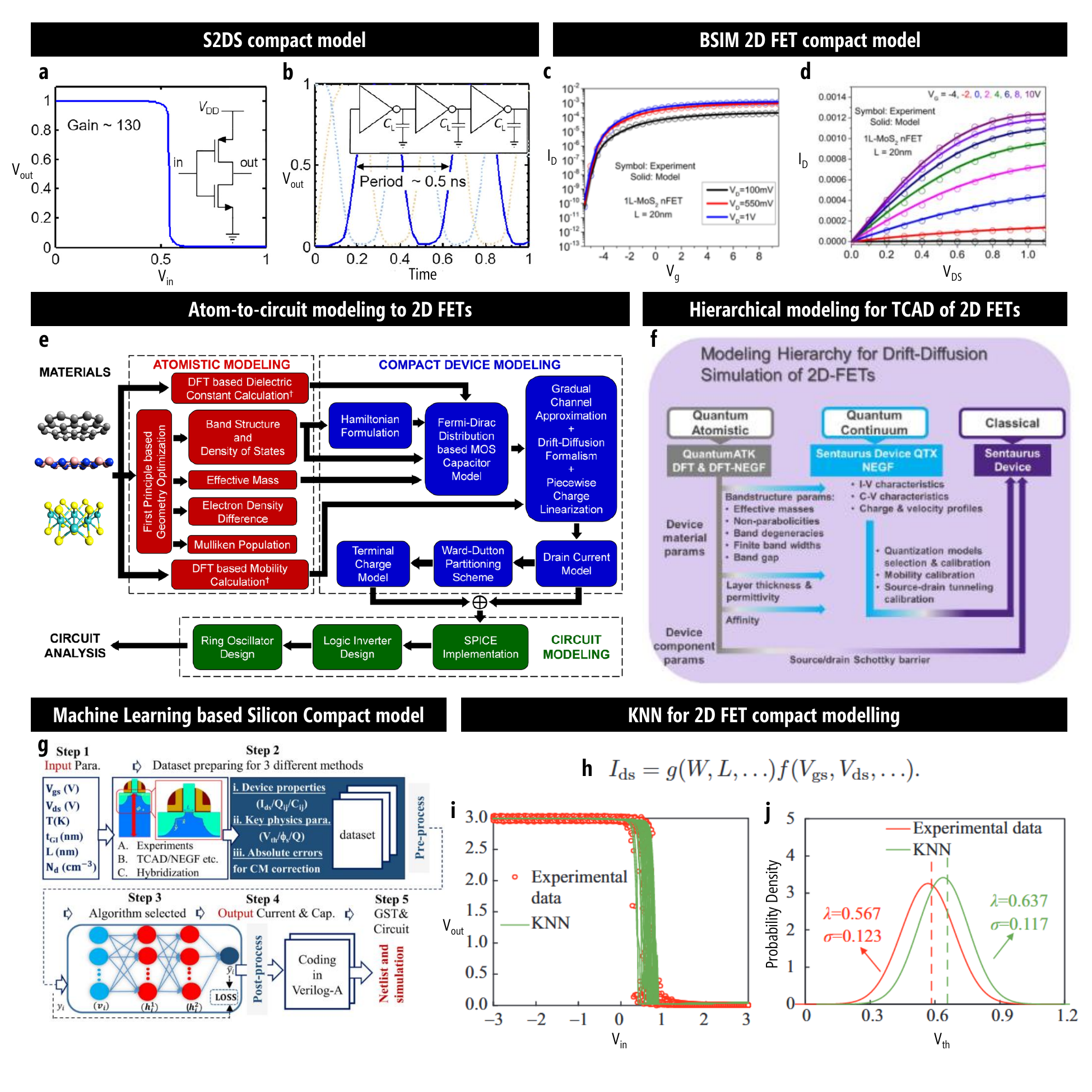}
    \caption{\textbf{Compact models for 2D semiconductor transistors} 
    (a, b) circuit simulation in S2DS compact model \cite{suryavanshiS2DSPhysicsbasedCompact2016}
    (a) Inverter sweep simulation
    (b) 3-stage oscillator simulation
    (c,d) mode validation  for a back-gated single-layer \ce{MoS2} FETs with self-heating in BSIM compact model of 2D FETs \cite{chenBSIMCompactModel2025}
    (c) $I_D$-$V_G$ plot
    (d) $I_D$-$V_D$ plot
    (e) Atom-to-circuit modeling to 2D FETs methodology \cite{dasAtomtocircuitModelingApproach2018}
    (f) Hierarchical modeling for TCAD of 2D FETs workflow \cite{silvestriHierarchicalModelingTCAD2023}
    (g) Procedure of ML compact model \cite{liOverviewEmergingSemiconductor2024}
    (h, i, j) Inverter validation in KNN for 2D FETs \cite{qiKnowledgebasedNeuralNetwork2023}
    (h) Drain current function
    (i) Experimental and simulation results
    (j) Threshold voltage statistic
    }
    \label{fig:compactmodel}
\end{figure*}

\subsection{Compact modelling of 2D semiconductors}

Compact models are implemented primarily through SPICE and Verilog-A. SPICE serves as the industry-standard circuit simulator, forming the computational backbone of CMOS design workflows \cite{tumaCircuitSimulationSPICE2009}. Verilog-A, in contrast, is an analog hardware description language that expresses device electrical behavior such as current–voltage relationships and charge dynamics in a form compatible with SPICE \cite{mijalkovicPracticalGuideVerilogA2022}. Modern compact models for 2D FETs are typically developed as SPICE-compatible Verilog-A codes, which enables their deployment in logic gate design, standard-cell development and circuit verifications \reftable{table:compact_model}. 

In this subsection, we focus on physics-based compact models for long-channel 2D transistors and multiscale compact models for short-channel 2D transistors, as well as data-driven compact models incorporating neural networks. We then highlight recent efforts in the performance projections of 2D circutis based on 2D compact modelling tools.

\begin{table*}[]
\begin{tabular}{|l|L{2cm}|L{3.5cm}|L{3.5cm}|L{2.5cm}|L{2cm}|L{2cm}|}
\hline
  \textbf{Reference} &
  \textbf{Modeling Paradigm} &
  \textbf{Core Transport Framework} &
  \textbf{Experimental Data Calibration} &
  \textbf{Circuit Level Demostration} &
  \textbf{HDL Implementation} &
  \textbf{SPICE-compatible} \\ \hline \hline
\cite{caoCompactCurrentVoltage2014} &
  Semiclassical &
  DD including Interface Traps, Mobility Degradation, and Inefficient Doping &
  Monolayer \ce{MoS2} NFET &
  Not demonstrated &
  Not implemented &
  Not reported \\ \hline
\cite{suryavanshiS2DSPhysicsbasedCompact2016} &
  Semiclassical &
  DD including  band structure, quantum capacitance, velocity saturation, contact resistance, and self-heating effects &
  Experimental monolayer \ce{MoS2} NFET with $L_g$ = 8 nm to 3.2 \si{\micro\meter} and   monolayer \ce{WSe2} PFET with $L_g$  = 9.4   \si{\micro\meter} &
  CMOS Inverter and 3-stage RO &
  Verilog-A &
  Yes (code available on nanoHUB) \\ \hline
\cite{wangSurfacePotentialBased2018} &
  Semiclassical &
  Surface Potential-Based compact Model with disorder-induced transport transitions &
  Experimental monolayer \ce{WS2} NFET with $L_g$ = 1.5 \si{\micro\meter} and BP nanoribbons   FET with $L_g$ = 1 \si{\micro\meter} &
  5-stage RO &
  Verilog-A &
  Not explicitly reported \\ \hline
\cite{yadavChargeBasedModelingTransition2018} &
  Semiclassical &
  Charge-based DD including interface traps, ambipolar transport, and negative capacitance &
  Experimental \ce{MoS2} n-type NCFET, ambipolar \ce{MoS2} FET with $L_g$ = 3.3 \si{\micro\meter},   MoTe2 FET with $L_g$ = 1 \si{\micro\meter}, and \ce{WSe2} PFET with $L_g$ = 9.4 \si{\micro\meter} &
  Not demonstrated &
  Not implemented &
  Not reported \\ \hline
\cite{dasAtomtocircuitModelingApproach2018} &
  Multiscale (DFT to Compact model) &
  DD with bias-dependent diffusivity and piecewise charge linearization &
  No direct calibration (DFT-derived parameters) &
  Inverter and 15-stage RO &
  Verilog-AMS &
  Yes \\ \hline
\cite{pasadasLargesignalModel2DFETs2019} &
  Semiclassical &
  DD with Fermi-Dirac statistics and Ward-Dutton charge partition &
  Experimental monolayer \ce{MoS2} NFET with $L_g$ = 150 nm &
  3-stage RO &
  Verilog-A &
  Not explicitly stated \\ \hline
\cite{ahsanSPICECompactModel2021} &
  Semiclassical &
  Explicit charge-based DD using 2D DOS, Fermi–Dirac statistics, Lambert-W, and Halley correction &
  Experimental monolayer \ce{MoS2} NFET with $L_g$ = 5 \si{\micro\meter}, MoTe2 FET with $L_g$ =   1 \si{\micro\meter}, and BP FET with $L_g$ = 0.6 \si{\micro\meter} (MoTe2 and BP FET are from   literature) &
  Inverter &
  Verilog-A &
  Yes \\ \hline
\cite{mounirCompactIVModel2023} &
  Semiclassical &
  Unified Charge Control Model (UCCM) based on Lambert-W function &
  Experimental double-gated monolayer \ce{MoS2} NFET, double-gated monolayer   \ce{WSe2} PFET, and single-back-gated multilayer \ce{MoS2} NFET &
  Not demonstrated &
  Not implemented &
  Not reported \\ \hline
\cite{qiKnowledgebasedNeuralNetwork2023} &
  Data-driven &
  Hybrid physics-embedded neural network (no explicit transport equation) &
  Experimental monolayer \ce{MoS2} NFET with $L_g$ = 20 - 40 \si{\micro\meter} &
  Inverter, 5-stage RO, NAND, NOR, DFF, and half-adder &
  Verilog-A &
  Not explicitly stated \\ \hline
\cite{silvestriHierarchicalModelingTCAD2023} &
  Multiscale (DFT to TCAD) &
  continuum effective mass (EM) NEGF transport calibrated into quantum-corrected TCAD &
  Simulated monolayer \ce{MoS2} NFET with $L_g$ = 10 nm &
  Not demonstrated &
  Not implemented &
  No \\ \hline
\cite{chenBSIMCompactModel2025} &
  Semiclassical &
  Charge-based DD integrated within the BSIM-CMG framework &
  Experimental double-gated three-layers \ce{MoS2} NFET with $L_g$ = 10 - 20 nm,   back-gated monolayer \ce{MoS2} NFET with $L_g$ = 20 - 75 nm, and nanosheet \ce{MoS2} NFET   with $L_g$ = 60 nm &
  Not demonstrated &
  Verilog-A (via BSIM-CMG) &
  Yes \\ \hline
\end{tabular}
\caption{\textbf{Compact models for 2D field-effect transistors.}}
\label{table:compact_model}
\end{table*}

\subsubsection{Physics-based compact modelling: Semiclasscial approach}

Semiclassical compact models for 2D FETs are typically built on DD transport coupled with Poisson electrostatics \reftable{table:compact_model}. A key distinction from bulk silicon is that the channel charge is obtained from the two-dimensional density of states combined with Fermi-Dirac statistics, which introduces quantum-capacitance effects that strongly influence the electrostatics and current–voltage characteristics of 2D FETs \cite{jimenezDriftdiffusionModelSingle2012, caoCompactCurrentVoltage2014}. 

Early DD-based model is established based on a surface-potential-based DD model for monolayer \ce{MoS2} FETs, which explicitly includes the 2D density of states and its impact on quantum capacitance to obtain analytical expressions for surface potential and drain current \cite{jimenezDriftdiffusionModelSingle2012}. A compact current–voltage model that continuously covers linear, saturation and subthreshold operation is also developed, which further includes interface traps, mobility degradation and inefficient doping effects that are particularly relevant for 2D transistors \cite{caoCompactCurrentVoltage2014}.

As 2D circuits grew in complexity, more comprehensive physics-based models emerged. A notable example is the Stanford 2D Semiconductor (S2DS) transistor model built on semiclassical transport models \cite{suryavanshiS2DSPhysicsbasedCompact2016}. S2DS explicitly accounts for contact resistance, traps and impurities, quantum capacitance, fringing fields, high-field velocity saturation and self-heating effects, and is implemented as SPICE-compatible Verilog-A code. S2DS has been calibrated against MoS$_2$ FET experimental data, and has been successfully employed to design inverters and ring oscillators with high gain and sub-nanosecond oscillation periods, exhibiting good agreement to experiemntal device measurements \cite{suryavanshiS2DSPhysicsbasedCompact2016}.

Beyond conventional DD model, several compact models that incorporate additional physical mechanisms important for 2D channels have also been developed. Charge-based models that couple the Khalatnikov equation with DD model have been constructed to describe ferroelectric and negative-capacitance effects in \ce{MoS2} FETs, while also capturing the charge trapping and ambipolar behaviors \cite{yadavChargeBasedModelingTransition2018, ahsanSPICECompactModel2021}. Analytical charge solutions using Halley’s correction is also developed for charge-based SPICE-compatible implementations \cite{ahsanSPICECompactModel2021}. Compact models based on multiple-trapping-and-release (MTR) theory have been proposed to better capture disorder-induced transport and subthreshold conduction in 2D channels \cite{wangSurfacePotentialBased2018}, thus extending the applicability of semiclassical models into regimes where trap-assisted conduction becomes significant. These physically-enriched compact models reflect the diversity of scattering and trapping mechanisms present in 2D semiconductors, which critically affect the device characteristics and performance.

More recently, BSIM compact model for 2D semiconductor FETs based on the BSIM-CMG core has been developed, extending the multi-gate FinFET/nanosheet formalism to 2D channels \cite{chenBSIMCompactModel2025}. Built on the BSIM-CMG framework for multi-gated silicon devices, the model extends CMG formulations to accommodate monolayer FET electrostatics and thermal effects. Its validation across multiple device geometries, including back-gated single-layer \ce{MoS2} FETs down to 20 nm channel lengths, thus demonstrating that industry-standard compact model can be adapted for 2D semiconductor and paving an important step toward eventual inclusion of 2D devices in commercial PDK and large-scale EDA workflows.

\subsubsection{Physics-based compact modelling: Hybrid quantum-semiclassical approach}

While semiclassical compact models provide an essential abstraction for circuit simulation, their accuracy becomes limited as 2D transistors scale into the near-10 or sub-10-nm regime, where quantum confinement, source-drain direct tunneling and contact-induced effects dominate transport. To bridge this gap, several multiscale hierarchical modelling frameworks have been developed to incorporate quantum-mechanical accuracy at the material and device levels while retaining the computational efficiency needed for circuit design.

A representative bottom-up approach integrates three layers of modelling, i.e. material, device and circuit, into a unified workflow \cite{dasAtomtocircuitModelingApproach2018} \figsub{fig:compactmodel}{e}. At the material level, DFT is employed to extract fundamental parameters such as band structure, effective mass, mobility, and dielectric constants. These quantities are then used to parameterize semiclassical DD compact models written in Verilog-AMS, thus enabling circuit-level SPICE simulations that retain key atomistic features of the underlying 2D channel. This apporach provides a systematic route for migrating material-level insights into circuit-level abstractions.

To further improve quantum accuracy, especially for short-channel 2D FETs, a hierarchical quantum-semiclassical modelling methodology has been demonstrated using the Synopsys TCAD software suites \figsub{fig:compactmodel}{f} \cite{silvestriHierarchicalModelingTCAD2023}. In this framework, DFT is first used to characterize 2D semiconductor band structures and other electrical properties. The Schottky barrier height is extracted from the DFT calculations of metal/channel contact heterostructures, thus ensuring that the quantum mechanical interfacial interactions are more accurately captured. These material inputs are fed into an effective-mass non-equilibrium Green’s function (EM-NEGF) solver implemented in Sentaurus Device QTX, which captures quantum ballistic transport and contact injection with significantly lower computational cost than full-band DFT-NEGF approaches (e.g., QuantumATK). The EM-NEGF current–voltage characteristics then serve as a reference for calibrating a semiclassical Sentaurus Device TCAD model. Notably, the calibrated DD model reproduces EM-NEGF results with high fidelity despite its semiclassical core based on DD models. 

These hybrid workflows, combining DFT-level material modelling, quantum-accurate NEGF transport and calibrated DD device models, offer a practical route toward compact models that retain quantum-level fidelity while remaining compatible with SPICE and EDA environments.

\begin{figure*}[t]
    \includegraphics[scale=0.47]{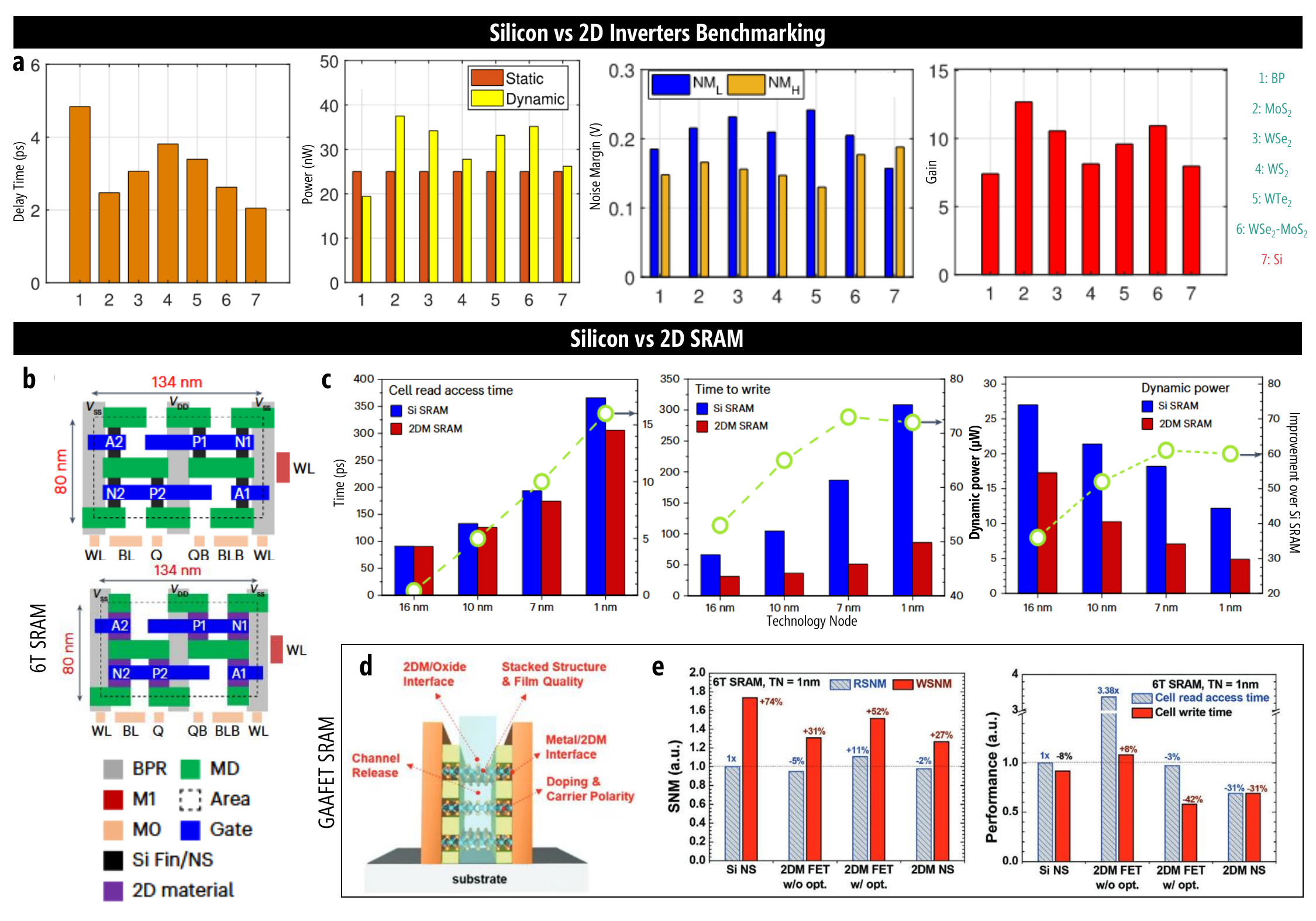}
    \caption{\textbf{Performance projection between silicon and 2D devices}
    (a) Bar charts of performance of 2D and Si inverters \cite{rawatPerformanceProjection2D2021}
    (b, c) Performance projection of Si and 2D SRAM \cite{luProjectedPerformanceSi2024}
    (b) Schematic of 6T SRAM
    (c) Si and 2D SRAM performance from various technology node
    (d, e) Performance projection of Si and 2D GAAFET SRAM \cite{liu2DMaterialsBasedStatic2022}
    (d) Illustration of GAAFET SRAM
    (e) Si, 2D SRAM, 2D SRAM with optimization, and 2D nanosheet SRAM performance}
    \label{fig:performance}
\end{figure*}

\subsubsection{Data-driven compact modelling}

The data-driven approach represents an emerging methodology for the compact modelling of 2D semiconductors. Instead of explicitly formulating device behavior from material physics and transport models, data-driven compact modelling uses machine-learning (ML) models trained on numerical simulation or experimental measurement data to reproduce the characteristics of a transistor. These models may take the form of look-up-table (LUT) representations or machine-learning–assisted compact models \cite{liOverviewEmergingSemiconductor2024} \figsub{fig:compactmodel}{g}. Similar techniques have recently gained traction in silicon compact modeling as well \cite{douPRIMEPhysicsRelatedIntelligent2025}, reflecting a broader shift toward AI-assisted device abstraction.

A representative example is the knowledge-based neural network (KNN) compact model for 2D FETs proposed by Qi \textit{et al.} \cite{qiKnowledgebasedNeuralNetwork2023}. In contrast to traditional neural network (TNN) approaches which treat all inputs such as geometry and bias as a single feature vector. The KNN framework partitions the input space into two physically meaningful groups: geometric parameters (e.g. channel length and width) and bias conditions (e.g., $V_\mathrm{GS}$, $V_\mathrm{DS}$). Two sub-networks are trained separately and then combined to predict the drain-source current \figsub{fig:compactmodel}{h}, thus improving interpretability and reducing overfitting relative to TNNs. The KNN model exhibits higher prediction accuracy for both long-channel (\SI{10}{\micro\meter}) and short-channel (\SI{0.08}{\micro\meter}) 2D MOSFETs. Its generality is demonstrated through applications to fabricated \ce{MoS2} circuits—including inverters, NAND, NOR, DFFs and half-adders, where the predicted transfer/output characteristics \figsub{fig:compactmodel}{i,j} and the oscillation frequency of a 5-stage \ce{MoS2} ring oscillator closely match experimental data. These results underscore the promise of data-driven approaches in developing scalable, SPICE-compatible compact models for 2D transistors.

Beyond direct current-voltage prediction, ML has also been used to accelerate quantum-transport simulations, offering a pathway to embed quantum-level physics directly into compact models. A bottom-up graph-featured neural network (GFGNN) that predicts the self-consistent potential in NEGF simulations of graphene-nanoribbon \textit{p–n} junctions and monolayer \ce{MoS2} MOSFETs has been recently developed \cite{xieBottomUpMachineLearningApproach2025}. When used to initialize the NEGF simulations, the ML model yields a simulation speed-up of approximately 1.7 to 4.2$\times$. When bypassing the self-consistent loop entirely, acceleration reaches 10 to 13$\times$, while maintaining millielectronvolt-level accuracy and excellent agreement in the current-voltage characteristics. More recently, a physics-knowledge-integrated hypervector neural network (PHVNN) trained on DFT–NEGF datasets of sub-5-nm gate-all-around FETs is introduced \cite{zhangPhysicsknowledgeintegratedNeuralNetwork2025}. By combining device-level geometric inputs with material descriptors such as band gap, effective mass, and Fermi-level shift, PHVNN predicts full transfer curves with a mean absolute error of $\sim$0.39 in $\log I_\mathrm{D}$ and nanosecond-level inference time. These recent works indicate that ML-accelerated NEGF tools can drastically reduce quantum device simulation time while retaining high fidelity, thus suggesting a viable route to future compact models where AI surrogates serve as the bridge between atomistic transport and circuit-level SPICE simulation for 2D semiconductor technologies.

\subsection{TCAD and compact-model simulations for performance projection and benchmarking of 2D circuits}

TCAD simulations and compact models play a central role in benchmarking 2D semiconductor devices against advanced silicon technologies and in projecting their performance at future technology nodes. By providing physics-based estimates of delay, power, noise margins and stability, these tools enable a quantitative assessment of the opportunities and limitations of 2D materials in circuit applications.

Device-and circuit-level NEGF simulations have been used to compare 2D inverters, including \ce{MoS2}, \ce{WSe2}, \ce{WS2}, \ce{WTe2}, and black phosphorus (BP), with scaled silicon CMOS inverters \cite{rawatPerformanceProjection2D2021} \figsub{fig:performance}{a}. 2D inverters generally exhibit lower off-state leakage, larger noise margins and higher gain than silicon, thus reflecting the excellent electrostatic control offered by atomically thin channels. However, the switching speed of 2D inverters remains lower than that of advanced silicon nodes, and the dynamic power dissipation is typically higher, with BP being a notable exception that shows competitive switching energy relative to silicon. Monolithic 3D architectures using 2D semiconductors have also been evaluated through EM–NEGF simulations, most notably for vertically stacked \ce{WS2} CMOS CFETs \cite{palThreedimensionalTransistorsTwodimensional2024a}. These simulations show that 2D-channel CFETs can deliver over 50\% improvement in inverter delay while enabling a tenfold increase in integration density relative to silicon nanosheet GAAFETs. This enhancement arises from the atomically thin channel body, which suppresses electrostatic degradation in stacked configurations and allows aggressive vertical scaling without incurring short-channel penalties.

SRAM has also been a key focus of TCAD-enabled benchmarking, given its sensitivity to transistor performance and its importance for memory-intensive workloads. Mixed-mode TCAD simulations of 6T SRAM cells \cite{luProjectedPerformanceSi2024} \figsub{fig:performance}{b} indicate that 2D-semiconductor-based SRAMs can achieve improved timing characteristics compared to silicon counterparts. Read-access time improves with technology scaling, reaching up to 16\% improvement at the 1 nm node \figsub{fig:performance}{c}. Write time and dynamic power exhibit their greatest improvements, i.e. 72\% and 60\%, respectively, around the 7 nm node, thus suggesting that 2D materials offer strong write-efficiency benefits in the near-term scaling regime.

For nanosheet-based SRAM architectures, TCAD benchmarking of gate-all-around FET (GAAFET) SRAM cells \cite{liu2DMaterialsBasedStatic2022} \figsub{fig:performance}{d} shows that 2D nanosheet SRAMs achieve faster read and write operations than silicon nanosheet SRAMs \figsub{fig:performance}{e}, thus highlighting the advantages of ultrathin channels in minimizing short-channel and parasitic effects. The primary trade-off lies in static noise margin: 2D-based SRAMs tend to exhibit reduced write static noise margin (WSNM) compared to silicon designs, which underscores the need for optimized device engineering and cell topologies.

Overall, TCAD and compact-model-based performance projections reveal a complex landscape: 2D semiconductor offer clear benefits in leakage, noise margin and timing, but challenges remain in switching speed, dynamic power and noise robustness. TCAD and compact models are thus necessary to provide essential insights and guidance for device optimization and circuit-informed co-design as 2D semiconductor advances toward integration potentially at future technological nodes.

\section{Challenges and Opportunities of 2D Semiconductor Chips}

\begin{figure*}[t]
    \includegraphics[scale=0.425]{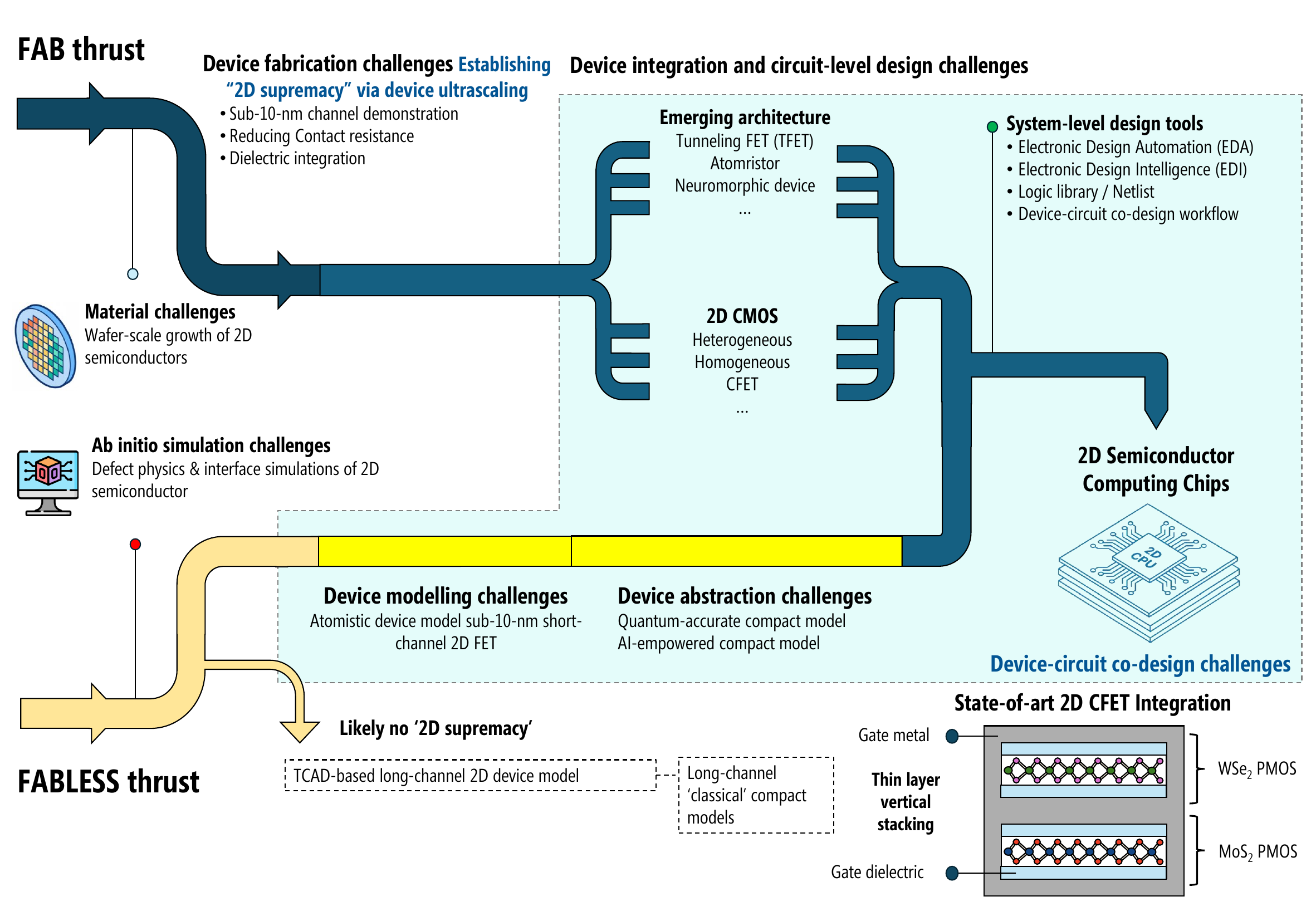}
    \caption{\textbf{Path and challenges towards 2D semiconductor computing chips} The schematic overview of 2D semiconductor technology for both FAB and FABLESS thrusts. Improving device fabrication in the FAB thrust supports establishing “2D supremacy” through superior ultrascaling beyond silicon technology. Novel and emerging architectures have been introduced, expanding the applications of 2D semiconductor technology. Especially for 2D CFETs, they have been fabricated utilizing very thin layers to stack 2D FETs as a single device. For the FABLESS thrust, TCAD-based classical computing for long-channel devices is an obstacle to achieving “2D supremacy.” Accurate quantum-based computing methods with lower computational cost are necessary for large-scale 2D circuit simulation. Bridging fabrication advances with quantum-accurate compact modeling and system-level design tools is critical to maximize the potential of 2D semiconductor technology.}
    \label{fig:challenge}
\end{figure*}

We now turn to the key challenges and opportunities in advancing 2D semiconductors from proof-of-concept circuits to practical chip-scale integration. While recent progress has established the feasibility of 2D transistors in logic, memory, and even microprocessor architectures, significant gaps remain in manufacturability, variability control, design–technology co-optimization, and ecosystem readiness. This section synthesizes the critical bottlenecks that currently limit large-scale adoption of 2D semiconductors, spanning both "FAB" and "FABLESS" thrusts \fig{fig:challenge}: materials synthesis, device integration, circuit robustness, and EDA infrastructure. The section also outlines emerging directions that could enable 2D technologies to transition from exploratory research to viable computing platforms.

\subsection{Where 2D semiconductors will and will not compete with silicon}
\label{subsec:where_2D_competes}

Before discussing specific technical challenges, it is worth identifying with intellectual honesty where 2D semiconductors are most and least likely to displace silicon. Such an assessment helps direct research effort toward applications where 2D technology offers a genuine advantage, rather than chasing parity with mature silicon in domains where the latter's accumulated optimization is essentially insurmountable.

\textbf{Domains where 2D semiconductors are unlikely to displace silicon.} For the highest-performance digital cores, where clock frequencies in the gigahertz range and aggressive power--performance optimization are required, 2D transistors are unlikely to surpass advanced silicon FinFETs or GAAFETs in the foreseeable future. Saturation drift velocities in monolayer TMDs are comparable to or below those of silicon, and contact resistance in 2D devices remains substantially higher than in scaled Si CMOS~\cite{choiRecentProgress1D2022, maVanWaalsContact2024, duScaledCrystallineAntimony2025}. Furthermore, the decades of investment in silicon design tools, compact models, PDKs, and process maturity create a substantial barrier to displacement in mainstream digital logic. Similarly, for analog and RF applications requiring high transconductance linearity and well-characterized noise behavior, silicon (and SiGe/III--V technologies) currently retain decisive advantages.

\textbf{Domains where 2D semiconductors offer genuine competitive advantages.} The strengths of 2D semiconductors lie in regimes where silicon's bulk physics breaks down or where its processing constraints are limiting. Four such domains stand out: (i) ultrascaled gate lengths below $\sim$5~nm, where silicon's electrostatic control degrades sharply but atomically thin 2D channels maintain robust gate coupling~\cite{meenaSub5Nm2D2023, yoonEnablingAngstromEra2025}; (ii) back-end-of-line monolithic 3D integration, where the low thermal budget of 2D synthesis allows direct integration on completed CMOS wafers~\cite{jayachandranThreedimensionalIntegrationTwodimensional2024, guoVanWaalsPolarityengineered2024, chowdhury3DIntegrationFunctionally2025}; (iii) flexible, wearable, and large-area electronics, where mechanical compliance and low-temperature processing are essential and where silicon is fundamentally unsuited~\cite{liLargescaleFlexibleTransparent2020, pengMediumscaleFlexibleIntegrated2024, tangLowPowerFlexible2023}; and (iv) compute-in-memory and neuromorphic architectures, where the compatibility of 2D devices with non-volatile and analog switching mechanisms enables architectures that conventional silicon CMOS implements only with significant overhead~\cite{migliatomaregaLogicinmemoryBasedAtomically2020, doddaAllinoneBioinspiredLowpower2022, yinEmerging2DMemory2021, wangInmemoryComputingArchitecture2021}.

Thus, the most credible near-term role for 2D semiconductors is not as a wholesale replacement for silicon CMOS in mainstream digital logic, but as a complementary technology -- extending silicon's reach into domains it cannot occupy efficiently, and providing dedicated layers in heterogeneously integrated systems. The longer-term prospect of fully 2D general-purpose computing depends on whether the device-level "2D supremacy" benchmark (Section~\ref{subsec:2D_supremacy}) can be met simultaneously with the chip-scale milestones outlined in Section~\ref{subsec:milestones}.
 
\subsection{Challenge 1. Homogeneous complementary 2D semiconductor for CMOS logic}

Although NMOS circuits based on \ce{MoS2} \cite{radisavljevicIntegratedCircuitsLogic2011, yuDesignModelingFabrication2016, wachterMicroprocessorBasedTwodimensional2017, wangElectronicDevicesCircuits2019, liLargescaleFlexibleTransparent2020} can be used to realize functional 2D chips, complementary integration of $n$- and $p$-type transistors such as CMOS logic is ultimately required to achieve compact circuit designs with high performance and low power consumption. Unlike bulk silicon, where appropriate doping readily enables both $n$-and $p$-type devices within the same material system, such an approach is not directly transferable to 2D semiconductors. For example, \ce{MoS2} typically exhibits $n$-type behavior, whereas \ce{WSe2} is commonly $p$-type. As a result, CMOS implementations often rely on \emph{heterogeneous integration} of two different 2D semiconductors, introducing additional fabrication and process complexity \cite{tianRailtoRailMoS2Inverters2022}.

Even if the fabrication challenges associated with heterogeneous integration can be mitigated, the use of dissimilar channel materials with different electrical properties can lead to current imbalance between pull-up and pull-down transistors, thus necessitating tedious device and circuit co-design. Identifying or engineering high-performance $p$-type 2D semiconductors \cite{dasHighperformancePtypeFieldeffect2025} that can effectively complement $n$-type \ce{MoS2} therefore remains a critical challenge before the full potential of 2D CMOS chips can be realized. An alternative and potentially attractive route is the development of \emph{homogeneous} CMOS platforms based on a single, ambipolar or $np$-tunable 2D semiconductor. Recent demonstrations of $np$-type \ce{MoTe2} via substitutional doping provide an early example in this direction \cite{panPrecisePtypeNtype2024}. More broadly, achieving near-symmetric $n$- and $p$-type performance within a single 2D material or developing device architectures that enable high-performance $p$-type operation using already mature materials such as \ce{MoS2}, could offer a promising pathway toward homogeneous 2D CMOS circuits with reduced fabrication complexity.

\subsection{Challenge 2. Design and compact modelling of large-scale 2D circuit}

Commercially available, industrial-grade transistor simulators are highly optimized and extensively validated for silicon technologies. However, their applicability to 2D transistors, where the atomically thin body requires fundamentally different electrostatic treatment and where charge transport physics deviates substantially from bulk silicon, remains an open question. In particular, 2D transistors in the sub-10-nm regime demand quantum-accurate atomistic simulations, such as NEGF and DFT, which are computationally prohibitive for large-scale circuit design and system-level exploration \cite{knoblochModeling2DMaterialBased2023}.
This gap highlights the need for \emph{2D-specialized} compact models that can faithfully capture quantum and electrostatic effects while remaining compatible with circuit-level simulation. Although multiple compact modeling efforts for 2D transistors have been reported \cite{suryavanshiS2DSPhysicsbasedCompact2016, chenBSIMCompactModel2025, qiKnowledgebasedNeuralNetwork2023, dasAtomtocircuitModelingApproach2018, silvestriHierarchicalModelingTCAD2023}, a standardized, quantum-aware compact model tailored for 2D semiconductors and compatible with industry-standard environments such as Verilog-A and SPICE is still lacking.
Recent demonstrations of hierarchical quantum–semiclassical hybrid approaches, which combine first-principles simulations at the material and interface levels with reduced-cost transport formalisms, illustrate a promising route toward quantum-accurate compact modeling for 2D devices \cite{dasAtomtocircuitModelingApproach2018, silvestriHierarchicalModelingTCAD2023}. Nevertheless, systematic validation of these approaches, along with standardization and benchmarking against experimentally fabricated devices and circuits, is still required to establish a robust and reliable 2D compact modeling framework.
Importantly, such a model would serve as a key enabler for the forward development of 2D chips, playing a foundational role analogous to that of BSIM in the silicon chip technology.

Data-driven compact modelling \cite{liOverviewEmergingSemiconductor2024, qiKnowledgebasedNeuralNetwork2023} also offers a complementary route to address the tool-chain gap for ultrascaled 2D circuits. Rather than deriving fully analytical expressions, ML-based compact models learn the mapping from biases, geometry, and process descriptors to $I$–$V$ and (increasingly) $Q$–$V$ characteristics, and can be deployed either as look-up tables or neural-network-based surrogate models such as graph neural network \cite{yangGraphBasedCompactModel2024} and physics-informed neural network \cite{gaoPhysicsInformedGraphNeural2020} in SPICE/Verilog-A environments. Importantly, recent work has moved beyond purely black-box fitting by embedding physical structure into the model, for example by separating geometry scaling from bias dependence (knowledge-based architectures) \cite{qiKnowledgebasedNeuralNetwork2023} or by explicitly training on derivatives to improve smoothness and numerical robustness required by circuit solvers \cite{guo2023physics, woo2022machine}.

\subsection{Challenge 3. Completing the standard cell library of 2D semiconductors}

The ecosystem of standard cell library remains largely incomplete for 2D semiconductors. Although preliminary efforts have established majority of the Boolean logic gates \cite{aoRISCV32bitMicroprocessor2025, ghoshComplementaryTwodimensionalMaterialbased2025}, a more complete arsenal of advanced logical cell designs are essential for achieving ultimately compact circuit layouts in large-scale digital systems. 2D semiconductor arithmetic cells beyond half-adder and full-adder \cite{pengMediumscaleFlexibleIntegrated2024, aoRISCV32bitMicroprocessor2025}, such as carry generators, multipliers, and subtractors, remain scarce, despite their importance in supporting complex numerical computations. Furthermore, advanced sequential cells commonly found in commercial standard cell libraries, especially the scan flip-flop used for debugging and error detection, are also missing thus far. The development of a more complete standard cell library is thus critically required for enabling more compact and efficient 2D chip designs.

\begin{figure*}[t]
    \includegraphics[scale=0.58]{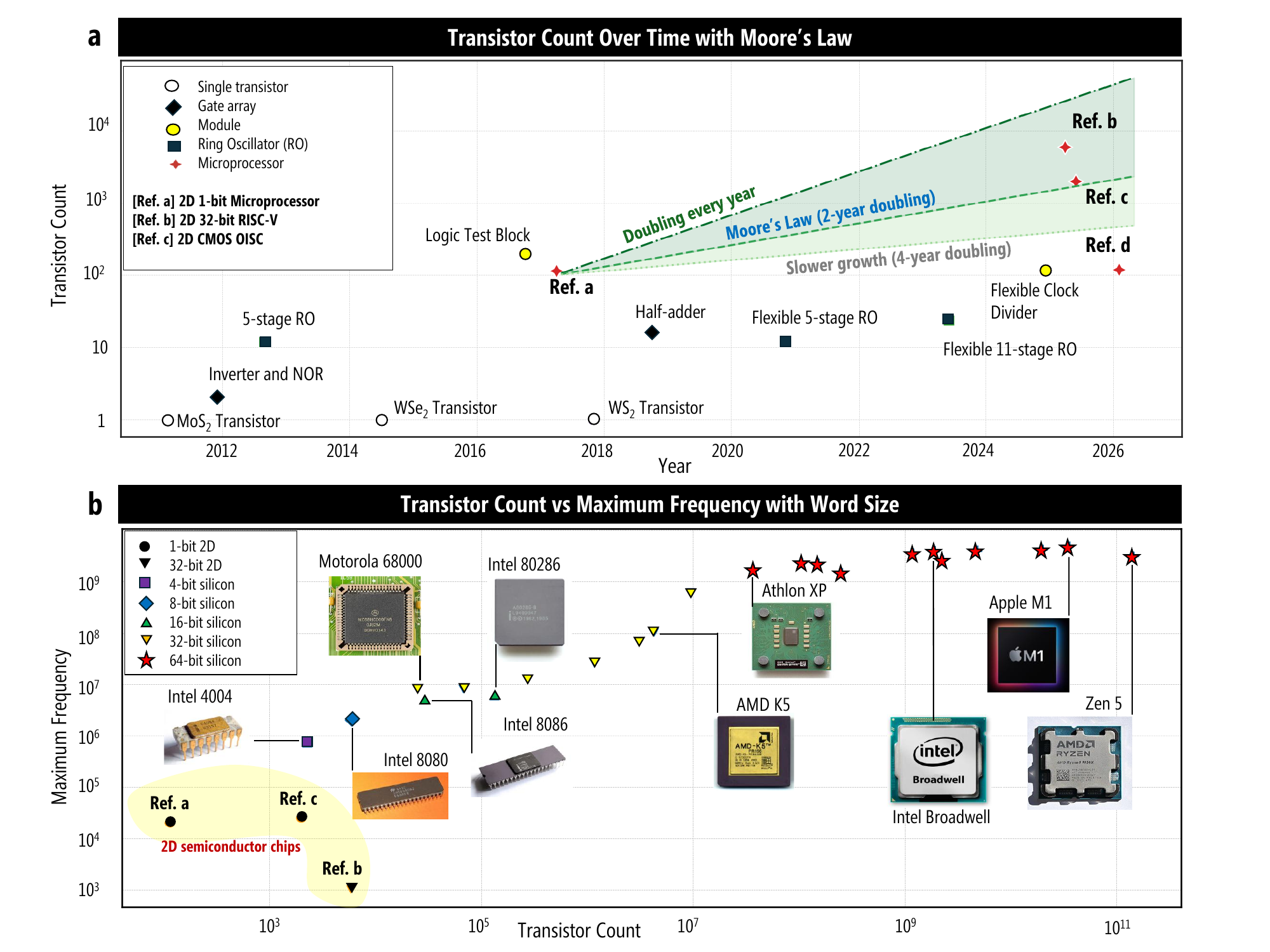}
    \caption{\textbf{Comparison between 2D semiconductor and silicon technology with Moore's law prediction} (a) Transistor count over time with Moore's law prediction for 2D semiconductor technology (b) Transistor count and maximum frequency with word size of 2D and silicon microprocessors plotting (Ref. a) 1-bit 2D microprocessor \cite{wachterMicroprocessorBasedTwodimensional2017} (Ref.b) RISC-V 2D microprocessor \cite{aoRISCV32bitMicroprocessor2025} (Ref.c) CMOS OISC 2D computer \cite{ghoshComplementaryTwodimensionalMaterialbased2025} (Ref.d) Towards-foundry 2D microprocessor \cite{guoTowardsfoundryStrategyCreating2026}}
    \label{fig:comparison}
\end{figure*}

\subsection{Challenge 4. Emerging transistor architectures for 2D chips}

Beyond conventional MOSFETs, 2D semiconductors can be incorporated into various emerging transistor architectures, such as tunnel FETs (TFETs), in which emphasize band-to-band tunneling to achieve a steep-slope characteristic \cite{kanungo2DMaterialsbasedNanoscale2022}. Cold-source FETs (CS-FETs) which utilize cold carrier injection contacts to suppress high-energy carriers, thereby reducing leakage current and improving the subthreshold slope \cite{yinComputationalStudyTransition2022}, offer another avenue towards ultralow power 2D devices. Moreover, 2D semiconductors has also been actively explored for their neuromorphic computing applications \cite{doddaAllinoneBioinspiredLowpower2022} such as the recent demonstration of large-scale implementation of MoS$_2$ memtransistors \cite{schranghamerLargescaleCrossbarArrays2025, chowdhury3DIntegrationFunctionally2025} These emerging 2D device highlight alternative pathways to integrate 2D semiconductors into future chip design beyond the conventional silicon CMOS paradigm.

\subsection{Challenge 5. Fabrication Yield and Device Ultrascaling Challenges: Demonstration of '2D supremacy'}

While 2D inverters can exhibit high fabrication yields, reaching around 99\% \cite{aoRISCV32bitMicroprocessor2025}, the overall yield of 2D chip fabrication suffers due to accumulated failures and the lack of fault-tolerant chip design\cite{wachterMicroprocessorBasedTwodimensional2017}. At the material and device levels, fabrication processes that minimize defects are critically needed for improvement \cite{wangCriticalChallengesDevelopment2024}. More critically, at the system level, fault-tolerant chip design should also form a crucial principle for ensuring the reliability of complex circuits and failure-sensitive chips \cite{wachterMicroprocessorBasedTwodimensional2017}.
For silicon, many design techniques have been adopted to overcome yield challenges. A common approach is implementing redundant multi-core processing units \cite{venkateshaSurveyFaultMitigation2021}. We thus expect a similar strategy to be tremendously useful in overcoming the yield challenge of 2D chips, especially in supplementing the material-and device-level limitations of 2D semiconductors.

Finally, we highlight another challenge of 2D semiconductor transistors, namely the ultrascaling of 2D smeiocnductors into the sub-10-nm channel length regime \cite{sakibHighperformanceMolybdenumDisulfide2025, duScaledCrystallineAntimony2025}. Majority of the 2D circuits are implemented using long-channel devices \fig{fig:celltable}. This stands in contrast to the key advantages of 2D semiconductors, which is to offer a viable route to suppress SCE in the sub-10-nm gate-length regime unachievable using silicon \cite{wangRoad2DSemiconductors2022, shenTrend2DTransistors2022}.
Overcoming the fabrication and device design challenges of sub-10-nm gate-length 2D semiconductor transistors, especially moving away from single `hero device' demonstration to circuit-level integration with good fabrication yield and device performance uniformity, shall be a key frontier of 2D semiconductor device research in the coming years -- a critical step to firmly establish the `2D supremacy' over silicon-based transistor in future technology nodes. 

\section{Future Prospects of 2D Semiconductor Chips}

\subsection{Gordon in the \emph{Flatland}: Moore’s Law Benchmarking of 2D Semiconductors}

While Moore’s law no longer governs the modern transistor scaling \cite{shalfFutureComputingMoores2020}, it continues to provide a useful lens in evaluating how quickly 2D semiconductor chips are closing the gap between device demonstrations and system-level integration.
Here we provide a benchmarking of 2D semiconductors against the Moore's law based on the three demonstration of 2D microprocessors thus far. The 2D microprocessors transistor count exhibits an encouraging growth trend that is comparable to Moore’s law projection \figsub{fig:comparison}{a}. In 2017, the field began with 115 transistors \cite{wachterMicroprocessorBasedTwodimensional2017}. If transistor count doubled every two years based Moore’s law, the expected transistor count for a 2D microprocessor in 2025 would be approximately 1,840. Both RISC-V microprocessor \cite{aoRISCV32bitMicroprocessor2025} and CMOS computer \cite{ghoshComplementaryTwodimensionalMaterialbased2025} have exceeded such projection, although we note that more 2D microprocessor realizations are needed before a more concrete projection can be established. Although the towards-foundry 2D \cite{guoTowardsfoundryStrategyCreating2026} chip does not exceed the expected transistor count, it introduces novel circuit design and fabrication methods for 2D semiconductor technology. Importantly, the effective adaptation of silicon digital circuit knowledge to the 2D semiconductor chips will be a key to accelerate the growth of 2D chips complexity and functionality.

We further project the performance of 2D microprocessors based on transistor count and operating frequency \figsub{fig:comparison}{b} as compared with silicon chips. Currently, 2D microprocessors have transistor counts comparable to those of first-generation commercial programmable silicon chips, such as the Intel 4004. However, their operating frequency is approximately 100 times lower than that of their silicon counterparts. The continual improvement of 2D transistors, especially in improving their switching speed is thus an urgent quest. 

In summary, when compared to the state-of-the-art commercial silicon microprocessors, 2D microprocessors still have a long way to go. Today’s silicon chips have reached beyond several billion transistors and operate at billions of cycles per second, featuring highly complex architectures. In contrast, the current state of 2D microprocessor technology is roughly equivalent to the early stage of silicon technology. However, the current demonstration of 32-bit RISC-V 2D microprocessors have already exceeded, in terms of word size, that of the first-generation silicon chips such as the 4-bit operation of early silicon chips \figsub{fig:comparison}{b}. This highlights the potential of 2D chips to advance more rapidly than silicon chips did in their initial stages of development.

\begin{figure*}[t]
    \includegraphics[scale=0.58]{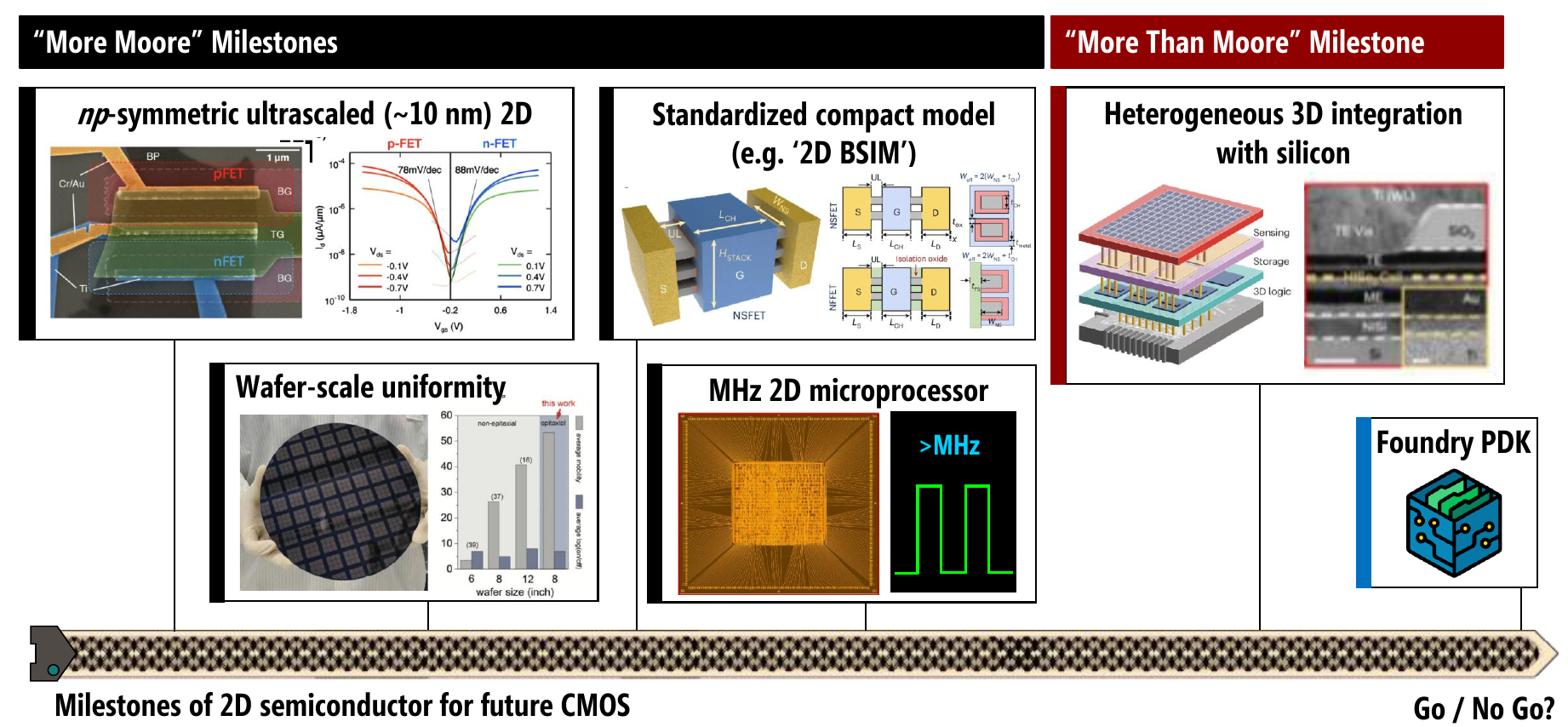}
    \caption{\textbf{Milestone Visualization} The roadmap consists five ``More Moore'' milestones and one ``More Than Moore'' milestone that must be achieved to establish 2D semiconductors as a credible technology for future computing.
    \textbf{(1) \textit{np}-symmetric ultrascaled ($\sim$10 nm) 2D FETs:} Simultaneous demonstration of n- and p-type 2D FETs at physical gate lengths at ultrashort channels with matched I–-V characteristics \cite{wuCMOSDevicesTwodimensional2019}.
    \textbf{(2) Wafer-scale uniformity:} A wafer hosting at least 10,000 functional sub-100-nm 2D transistors with threshold-voltage variability, validating statistical uniformity required for circuit-grade integration \cite{yuEightWaferScaleEpitaxial2024}.
    \textbf{(3) Standardized compact model (e.g., `2D BSIM'):} Release of a SPICE-compatible compact model mirror the foundational role of BSIM for silicon \cite{palThreedimensionalTransistorsTwodimensional2024a}.
    \textbf{(4) MHz 2D microprocessor:} A 2D microprocessor operating above 10 MHz that executing a non-trivial multi-bit arithmetic instruction sequence, closing the $\sim$$10^{5}\times$ frequency gap relative to first-generation silicon chips at comparable transistor counts \cite{aoRISCV32bitMicroprocessor2025}.
    \textbf{(5) Heterogeneous 3D integration with silicon:} Monolithic integration of a 2D semiconductor logic block on a completed CMOS Si substrate via back-end-of-line processing \cite{jayachandranThreedimensionalIntegrationTwodimensional2024, jainHeterogeneousIntegration2D2025}.
    \textbf{Foundry PDK:} This step mark the transition from academic demonstration to industrially accessible technology. Failure to achieve any of these milestones by the 2030s would indicate that current bottlenecks are more fundamental than presently anticipated.}
    \label{fig:milestone}
\end{figure*}

\subsection{Near-term key milestones of 2D semiconductor computing chips}
\label{subsec:milestones}

Establishing a clear set of measurable milestones is essential to gauge whether 2D semiconductor chips are progressing from exploratory research toward technological readiness. While precise timelines depend on coordinated advances across materials, devices, modeling and EDA infrastructure, we propose the following five concrete milestones as benchmarks against which future progress can be assessed. These targets are deliberately specific and falsifiable, in contrast to broader aspirational statements, and are intended to provide a shared reference for both the device and circuit communities. They are not arbitrary aspirations but concrete benchmarks within the application space identified in Section~\ref{subsec:where_2D_competes}, i.e. targeting domains where 2D semiconductors offer a genuine competitive advantage over silicon, rather than seeking parity with silicon in domains it has already optimized.

\emph{Milestone 1 --- Symmetric $n$-FET and $p$-FET performance at ultrascaled gate length.} A simultaneous demonstration of $n$-type and $p$-type 2D FETs at physical gate length below 12~nm with matched on-current and contact resistance within a factor of $\sim$2 of each other. This milestone targets the intersection of two of the field's most acute bottlenecks. On the n-side, monolayer \ce{MoS2} FETs with semimetal contacts (Bi and Sb) have reached contact resistance as low as $\sim$123~$\Omega\cdot\mu$m for Bi \cite{shenUltralowContactResistance2021} and $\sim$660~$\Omega\cdot\mu$m for the more thermally stable Sb \cite{liApproachingQuantumLimit2023,chouAntimonySemimetalContact2021}, approaching the quantum limit and enabling on-currents above 1~mA/$\mu$m. The $p$-type 2D FET, however, lags by roughly an order of magnitude: state-of-the-art \ce{WSe2} $p$-FETs with degenerate doping achieve $R_c \sim 230$--$642$~$\Omega\cdot\mu$m only under specialized doping schemes such as \ce{PtCl4} or tungsten oxyselenide monolayers~\cite{kimLowContactResistance2024, borahLowResistancePTypeOhmic2021}, while doping-free vdW $p$-contacts on \ce{WSe2} typically remain in the k$\Omega\cdot\mu$m regime~\cite{xieLowResistanceContact2024}. Closing this asymmetry is a prerequisite for balanced 2D CMOS, since unequal $n$-/$p$-side drive strength directly degrades inverter noise margins and compromises CMOS power-efficiency advantages. Achieving this milestone at gate length below 12~nm -- the projected stopping point for silicon gate-length scaling per the IRDS roadmap~\cite{IRDS2024IRDS}---would constitute the first concrete demonstration of "2D supremacy" (Section~\ref{subsec:2D_supremacy}), since it operates in a regime that bulk silicon physics cannot reach. While individual sub-10-nm $n$-type 2D demonstrations have been reported~\cite{desaiMoS2Transistors1nanometer2016, sakibHighperformanceMolybdenumDisulfide2025}, no comparable p-side demonstration exists, and no demonstration to date pairs the two within a single fabrication flow.

\emph{Milestone 2 -- Wafer-scale uniformity.} A demonstration of a wafer with at least 10{,}000 functional 2D transistors at sub-100-nm gate length, with measured threshold-voltage variability $\sigma V_{\mathrm{th}} < 50$~mV across the wafer and a fabrication yield exceeding 95\%. This bar tests whether the recent breakthroughs in wafer-scale \ce{MoS2} growth~\cite{liu2026kinetic, xia12inchGrowthUniform2023} can sustain the statistical uniformity required for circuit-grade integration. Milestone~1 in particular operationalizes the device-level criterion (v) of the "2D supremacy" benchmark (see Section~\ref{subsec:2D_supremacy}).

\emph{Milestone 3 -- Standardized compact model.} The release of a SPICE-compatible compact model for a 2D semiconductor FET that has been independently validated by at least two research groups against experimental sub-10-nm devices, and adopted within at least one publicly available PDK. This milestone would mirror the role that BSIM played for silicon and is essential for translating device physics into scalable circuit design~\cite{chenBSIMCompactModel2025, knoblochModeling2DMaterialBased2023}.

\emph{Milestone 4 -- Functional 2D microprocessor at megahertz operation.} A 2D microprocessor with measured clock frequency exceeding 10~MHz, operating at supply voltage below 2~V, and demonstrating execution of a non-trivial instruction sequence (e.g. a multi-bit arithmetic benchmark). Current 2D microprocessor demonstrations~\cite{aoRISCV32bitMicroprocessor2025, wachterMicroprocessorBasedTwodimensional2017, ghoshComplementaryTwodimensionalMaterialbased2025, guoTowardsfoundryStrategyCreating2026} operate at kHz frequencies, approximately $10^5\times$ slower than first-generation silicon chips at comparable transistor counts. Closing this gap is critical for credibility of 2D semiconductors.

\emph{Milestone 5 -- Monolithic heterogeneous 3D integration.} A demonstration of a 2D-semiconductor logic block monolithically integrated on a CMOS Si substrate (BEOL integration), with measured functionality of both layers and a quantitative assessment of cross-layer parasitics. This would establish 2D semiconductors as a complement to, rather than replacement for, silicon -- exploiting their thermal-budget advantage~\cite{jayachandranThreedimensionalIntegrationTwodimensional2024, guoVanWaalsPolarityengineered2024, chowdhury3DIntegrationFunctionally2025, jainHeterogeneousIntegration2D2025, ghoshMonolithicHeterogeneousThreedimensional2024} for hybrid systems.

\emph{Milestone 6 -- Foundry PDK release.} The release of a PDK for at least one species of 2D semiconductor transistor by a foundry or consortium, including design rules, standard cell library, compact models and parasitic extraction templates \cite{mao2DSPARK2025}. This milestone signals the transition from academic demonstration to industrially accessible technology -- a step that, for example, GaN and SiC technologies took roughly two decades after their first device demonstrations.

Achievement of these milestones within the immediate decade would indicate that 2D semiconductors are on a credible trajectory toward commercial relevance. Failure to achieve any of them by 2030s would suggest that the field's bottlenecks are more fundamental than presently anticipated.

\section{Conclusion}

2D semiconductors have emerged as promising candidates for extending computing beyond the physical limits of silicon in the post-Moore’s law era. Their atomically thin channels enable excellent electrostatic control, aggressive scalability, and energy-efficient operation. Research is rapidly advancing from early proof-of-concept devices, including \ce{MoS2} inverters and logic gates, to increasingly complex integrated systems such as RISC-V microprocessors \cite{aoRISCV32bitMicroprocessor2025} and CMOS OISC computers \cite{ghoshComplementaryTwodimensionalMaterialbased2025}. In parallel, advances in compact modeling and simulation frameworks are enabling practical large-scale 2D circuit design.

In the post-Moore’s scaling era, 2D semiconductors are well positioned to sustain continued device scaling while enabling new computing paradigms. Initially, these materials can complement silicon in specialized applications such as flexible electronics, neuromorphic computing, and edge devices. Ultimately, advances in material synthesis, fabrication, and integration will enable 2D semiconductors to support commercial general-purpose computing architectures at the angstrom scale. Realizing this opportunity requires close collaboration between material physicists and circuit designers to translate atomic-scale innovations into practical computing technologies. Through this interdisciplinary effort, 2D semiconductors are poised to define the next generation of the semiconductor industry.

\section{Acknowledgment}

This work is supported by the Singapore National Research Foundation (NRF) Frontier Science Competitive Research Programme (F-CRP) under the award number NRF-F-CRP-2024-0001. Y.S.A. acknowledges the supports from the Kwan Im Thong Hood Cho Temple Early Career Chair Professorship. J.L. acknowledges the supports from the Ministry of Science and Technology of China (No. 2022YFA1203904), the National Natural Science Foundation of China (No. 12274002). C.S.L. acknowledges the supports from the A*STAR under its MTC IRG (Grant No. M23M6c0103). 

\bibliography{ChipReview}

\end{document}